\documentclass[aps,preprint,showpacs,preprintnumbers,amsmath,amssymb]{article}
\usepackage{float}
\usepackage[pdftex]{graphicx}
\RequirePackage{graphicx}
\RequirePackage{flushend}
\RequirePackage[colorlinks,citecolor=blue,urlcolor=blue,linkcolor=blue]{hyperref}
\usepackage{slashed}
\usepackage{verbatim}
\usepackage{graphicx}
\usepackage{combelow} 
\usepackage{mathtools}
\usepackage[usenames,dvipsnames,svgnames,table]{xcolor}
\usepackage{soul}
\usepackage{ulem}
\usepackage{braket}
\usepackage{bm}
\usepackage{amssymb} 

\begin{document}

\title{Emission of $W$ bosons by electrons and decays of $W$ bosons in the early Universe}
\author{Amalia Dariana Fodor and Cosmin Crucean\\ \thanks{amalia.fodor97@e-uvt.ro, cosmin.crucean@e-uvt.ro}\\
	{\small \it Faculty of Physics, West University of Timi\c soara,}\\
	{\small \it V. Parvan Avenue 4 RO-300223 Timi\c soara,  Romania}}

\maketitle

\begin{abstract}
	In this paper we study the transition rates for the decay of $W^{\pm}$ bosons into leptons in de Sitter spacetime. From the results obtained in the de Sitter geometry we recover the Minkowski limits of the transition rates. We also study the emission of $W^{\pm}$ bosons by leptons, a process which is possible only in the large expansion conditions of the early Universe. These results are obtainable due to perturbative methods which allow us to compute the transition amplitudes in the first order of perturbation theory. The resulting probabilities and rates depend on the particle masses and the Hubble parameter. In order to establish finite expressions for the probabilities and transition rates, we make use of dimensional regularization and the minimal substraction method.
\end{abstract}

\section{\bf{Introduction}}

Massive $W$ bosons are known to be unstable and decay into leptons \cite{12,rat}. The rates for these decays were computed in the Minkowski electro-weak theory, and were measured and well studied \cite{12,rat}. For that reason, it is important to understand how these decay rates would be modified by a curved background. This is a fundamental issue, since the charged massive bosons $W^{\pm}$ could exist as stable particles only in the early Universe, when the large expansion and high temperatures made the background energy far greater than the rest energy of the $W$ bosons \cite{w1,w2}. As the temperatures dropped and the background energy decreased, these bosons decayed into stable particles, such as electrons and neutrinos. Thus, the previous results prove that, in the very early stages of the Universe, when the temperatures and densities were high enough, these bosons could exist in large numbers at thermal equilibrium \cite{w1,w2}. In this paper, we want to understand how the creation and decay rates for standard model particles were modified in the conditions of the early Universe, when the expansion rate was large. 

By using a perturbative approach for computing quantities of interest, such as the transition amplitudes and transition rates, we are close to the principles of quantum field theory in flat spacetime. The perturbative approach helps us establish a connection between the results from flat space theory and the more general results in the de Sitter geometry \cite{23,rc,24,26,27,28,31,b1,b2,cc,43,44,45,46,47}. Our results in the de Sitter geometry will depend on the Hubble parameter and will help us understand how the transition probabilities and rates change from the large expansion conditions of the early Universe to no expansion in the Minkowski limit, when the Hubble parameter is zero. 

In flat spacetime QFT, massive vector fields are described by the Proca equation, proposed long ago in \cite{PR1}. The Proca equation is fundamental for understanding the interactions between massive bosons and leptons in flat space field theory \cite{12,19,20,rat}. When one tries to develop the theory of massive vector fields on a curved background there are serious mathematical challenges that need to be addressed. For these reasons the Proca equation is less studied in curved spacetimes. However, a few fundamental results concerning the Proca field in curved spacetimes can be found in \cite{2,WT}. These results \cite{2,WT,38} concern the problem of fundamental solutions for the Proca equation in de Sitter spacetime, as well the propagators for the Proca field in curved spacetimes. The Dirac equation \cite{PC,22} and the fermionic propagator were also studied, in de Sitter spacetime and in more general Robertson-Walker metrics \cite{CRR,COT,22,rfv,rfv1}. At this moment we have important results in the theory of free fields and propagators in the de Sitter geometry \cite{15,17,18,PT,COT,WF,LL,LL1,29,30,32,40,41}. It is also important to study elementary processes that emerge from the field theory on curved spacetime, such as first order perturbative transitions that could generate particles. The results obtained so far prove that massive charged $W^{\pm}$ bosons could be generated in the early Universe \cite{w1}, using a perturbative approach \cite{44}. It is also a well established fact that the conditions of the early Universe are suitable for a perturbative treatment of the particle production problem. One of the approaches proposed in \cite{17,18} suggests that the particle production could also happen when quantum fields interact and extended studies regarding this method can be found in \cite{24,25,26,27,28,29,30,31,32,b1,b2,cc,33,34,35}. This method is based on the computations related to the first order processes in a curved background \cite{15,LL,LL1,23,24,25,26,27,28,29,30,31,32,b1,b2,33,34,35}, with the observation that these processes are forbidden in flat space field theory by the energy-momentum conservation laws. We mention that the theory electro-weak interactions in the de Sitter geometry, was proposed in \cite{cc,43,44,45,47}.  

In the present work, we use the results obtained in \cite{24,43,44}, where perturbative methods were used for defining the transition amplitudes in de Sitter spacetime. These results prove that the Weinberg-Salam theory of electro-weak interactions in Minkowski spacetime  \cite{PR1,w1,12,3,4,5,6,7,8,9,10,11,cr} can be obtained from a more general theory of weak interactions in the de Sitter geometry \cite{cc,43,44,45}. For developing the field theory in a curved background we use the minimal coupling of the Dirac and Proca fields to gravity. In the present paper we will also consider, for the first time, the problem of spontaneous emission of the $W^-$ boson and a neutrino, from an electron , $e^-\rightarrow W^-+\nu_e$, as well as the decays of the $W$ charged bosons, such as $W^- \rightarrow e^-+\widetilde{\nu}_e$. We also mention that the theory of interactions between the charged $W^{\pm}$ bosons and fermions were constructed in \cite{44} by using the Lagrangian densities for the free Dirac and Proca fields, together with the interaction Lagrangian density, constructed with the help of the charged currents. In \cite{cc,43,44,45}, it was proven that the massive $W,Z$ bosons were generated in the very early Universe when the expansion parameter was larger than their masses. 

It is also important to mention non-perturbative approach related to the problem of massive boson generation obtained in \cite{37}. Our approach does not contradict the result in \cite{37}, and should be seen as a completion of the picture related to the problem of particle production in the early Universe. The fundamental results obtained in \cite{32,33,34,35}, introduce for the first time the idea that the spacetime expansion could generate particles. More recently, the phenomenon of particle production using perturbative methods was extensively presented in  \cite{24,25,26,27,28,29,30,31,32,b1,b2,cc,33,34,35,43,44,45,46,47}. The first papers related to the problem of particle production using the perturbative approach can be found in \cite{17,18}.

This paper is organized as follows: in the second section we present the decay of $W$ into leptons in de Sitter spacetime. We compute the transition amplitude, as well as the total transition rate in the limit of large expansion. Afterwards we move to compute the total transition rate in the Minkowski limit, and recover the well known rate of this process in Minkowski spacetime. The third section is dedicated to the problem of $W$ boson emission by electrons in the de Sitter geometry. After we compute the transition amplitude, we also find out the total transition rate of this process in the general case. Our graphical analysis shows that this rate vanishes in the Minkowski limit, when the expansion factor is zero, as it should. The total probabilities and transition rates are evaluated throughout this paper by using dimensional regularization and the subtraction method. The conclusions of our study are summarized in section four. The \textbf{Appendix} contains sections \textbf{A} and \textbf{B}. In \hyperref[AppendixA]{Appendix A} we give different useful formulas which we use throughout this paper. The main steps for obtaining the Minkowski limit from our amplitude are given in \hyperref[AppendixB]{Appendix B}. This result helps us establish the final form of the transition rate in the Minkowski limit.

\section{\bf{Leptonic decay of $W^{\pm}$ bosons}}

The transition amplitudes of the electro-weak theory in de Sitter geometry were defined in \cite{43,44} and we will use these results in our present computations. Our study will be done on a de Sitter spacetime, with the line element given in conformal form \cite{1}:
\begin{equation}\label{metr}
ds^2=dt^2-e^{2\omega t}d\vec{x}^2=\frac{1}{(\omega t_{c})^2}(dt_{c}^2-d\vec{x}^2)
\end{equation}

In the above equation the conformal time is defined in terms of the proper time by $t_{c}=-e^{-\omega t}/\omega$, where $\omega$ is the Hubble parameter ($\omega>0$).

Our computations will be done in the de Sitter geometry, where we consider the conformal chart with conformal time $t_{c}\in(-\infty,0)$, which covers the expanding portion of de Sitter spacetime \cite{1}. For de Sitter line element in the Cartesian gauge (\ref{metr}), we have the non-vanishing tetrad components:
\begin{equation}
e^{0}_{\widehat{0}}=-\omega t_c  ;\,\,\,e^{i}_{\widehat{j}}=-\delta^{i}_{\widehat{j}}\,\omega t_c.
\end{equation}

The leptonic decay of massive charged bosons can happen in six ways, three for each charged boson:
\begin{align}
&W^{-} \rightarrow e^{-} + \tilde{\nu}_{e}  \quad\quad\quad\quad\,\quad  W^{+} \rightarrow e^{+} + \nu_e \nonumber\\
&W^{-} \rightarrow \mu^{-} + \tilde{\nu}_{\mu}  \quad\quad\quad\quad\quad  W^{+} \rightarrow \mu^{+} + \nu_\mu \nonumber\\
&W^{-} \rightarrow \tau^{-} + \tilde{\nu}_{\tau}  \quad\quad\quad\quad\,\quad  W^{+} \rightarrow \tau^{+} + \nu_\tau \nonumber
\end{align}

The process we are going to use as an example is the decay of a negatively charged $W$ boson:
\begin{equation}
W^{-} \rightarrow e^{-} + \tilde{\nu_e}.
\end{equation}

In this case the transition amplitude is the following \cite{44}:
\begin{align}\label{AmpDecayW}
\mathcal{A}_{[{W^{-} \rightarrow e^{-} + \tilde{\nu}}]} &= \frac{i g}{2\sqrt{2}} \int d^4x \sqrt{-g(x)}\bar{u}_{\vec{p}\sigma}(x) \gamma^{\hat{\alpha}} e^{\mu}_{\hat{\alpha}}(1-\gamma^5) v_{\vec{p' \sigma'}}(x) f_{\vec{P}\lambda,_{\mu}}(x),
\end{align}
where $v_{\vec{p' \sigma'}}(x)$, $f_{\vec{P}\lambda,_{\mu}}(x)$ and $u_{\vec{p}\sigma}(x)$ are the the field equation solutions for the antineutrino, the $W^{-}$ boson and the electron, respectively, keeping in mind that $\bar{u} = u^{\dag} \gamma^{0}$.

In our computations we use the chiral representation of the Dirac matrices \cite{19,20}:
\begin{align}
\gamma^{0}=
\begin{pmatrix}
0\quad & \,\,\,1\\
1\quad & \,\,\,0
\end{pmatrix}\quad
\gamma^{i}=
\begin{pmatrix}
0 & \sigma^{i}\\
-\sigma^{i} & 0
\end{pmatrix}\quad
\gamma^{5}=
\begin{pmatrix}
-1 & \,\,\,0\\
0 & \,\,\,1
\end{pmatrix}, 
\end{align}
where $\sigma^{i}$ are the Pauli matrices and $1$ represents the identity matrix $I_2$.

The solutions for the Proca field in the de Sitter spacetime, which describe massive bosonic fields, were computed in \cite{2}. The solutions
of the Proca field depend on the conformal time $t_c$, the momentum $\vec{P}$ and polarisation $\lambda$. They also differ with respect to the values of polarisation $\lambda$, which can be 0 or ±1 \cite{2}. 

For $\lambda=\pm 1$ the temporal component vanishes and we remain with the spatial component:
\begin{align}\label{W:solProcaLambda1}
f_{\vec{P}\lambda=\pm 1,_i}(x) &= 
\frac{\sqrt{\pi}e^{-\pi K/2}\sqrt{-t_c}}{2(2\pi)^{3/2}} H_{iK}^{(1)}(-Pt_c) \epsilon_i(\vec{P},\lambda=\pm 1)e^{i\vec{P}\vec{x}},
\end{align}
while for $\lambda = 0$ we have both temporal:
\begin{align}\label{W:solProcaLambda0t}
f_{\vec{P}\lambda=0,_0}(x) &= 
\frac{\sqrt{\pi}e^{-\pi K/2}\omega P (-t_c)^{3/2}}{2(2\pi)^{3/2}M_W} H_{iK}^{(1)}(-Pt_c) e^{i\vec{P}\vec{x}}
\end{align}
and spatial components \cite{2}:
\begin{align}\label{W:solProcaLambda0s}
f_{\vec{P}\lambda=0,_i}(x) &=
\frac{i\sqrt{\pi}e^{-\pi K/2}\omega P}{2(2\pi)^{3/2}M_W}\bigg[\left(\frac{1}{2}+iK\right)\frac{\sqrt{-t_c}}{P} H_{iK}^{(1)}(-Pt_c)- (-t_c)^{3/2} H_{1+iK}^{(1)}(-Pt_c)\bigg]\nonumber\\
&\times\epsilon_i(\vec{P},\lambda= 0) e^{i\vec{P}\vec{x}},
\end{align}
where $\epsilon_i(\vec{K},\lambda)$ are the polarization vectors, $H^{(1)}(z)$ are Hankel functions of the first kind and the index is defined by $K=\sqrt{\left(M_W/\omega\right)^2-1/4}$. By $M_W$ we denote the mass of the $W^{\pm}$ bosons, which is $M_W \approx 80$ GeV, while $\omega$ is the Hubble parameter.

For the fermions we will make use of the solutions of the massive and massless Dirac fields in the de Sitter geometry \cite{35}. The electron of momentum $\vec{p}$ and helicity $\sigma=\pm 1/2$ is described by solution \cite{35}:
\begin{align}\label{W:solelectron}
u_{\vec{p},\sigma}^{\dag}(t,\vec{x})&=\frac{-i}{(2\pi)^{3/2}}\sqrt{\frac{\pi p}{\omega}}
\begin{pmatrix}
\frac{1}{2}e^{k\frac{\pi}{2}}H_{\nu_{+}}^{(2)}(\frac{p}{\omega}e^{-\omega t})\xi_{\sigma}^{\dag}(\vec{p}), \,\,\,
\sigma e^{-k\frac{\pi}{2}} H_{\nu_{-}}^{(2)}(\frac{p}{\omega}e^{-\omega t})\xi_{\sigma}^{\dag}(\vec{p}) \\
\end{pmatrix}
\nonumber\\
&\times e^{-i\vec{p}\vec{x}-2\omega t},
\end{align}
which is given with respect to the proper time $t$ and helicity spinors $\xi_{\sigma}(\vec{p})$. The order of the Hankel functions of the second kind $H^{(2)}$ is $\nu_{\pm}=1/2 \pm ik$, where $k=m_e/\omega$ is the ratio between the electron mass $m_e$ and the Hubble parameter. 

The antineutrino solution with respect to the conformal time $t_c$ and helicity spinors $\eta_{\sigma'}(\vec{p'})$ is \cite{35}:
\begin{align}\label{W:solantineutrino}	
v_{\vec{p'},\sigma'}(t_{c},\vec{x})
&=  \left(\frac{-\omega t_{c}}{2\pi}\right)^{3/2} \left(
\begin{array}{c}
\left(\frac{1}{2}+\sigma' \right)\eta_{\sigma'}(\vec{p'}) \\
0 \\
\end{array}\right) e^{ip' t_{c}-i\vec{p'}\vec{x}},
\end{align}
where $\vec{p'}$ and $\sigma'=\pm 1/2$ are the antineutrino momentum and helicity, respectively.

\subsection{\bf{ The amplitude for $\lambda = \pm 1$ }}

The amplitude for $\lambda = \pm 1$ can be obtained by replacing solutions \eqref{W:solProcaLambda1}, \eqref{W:solelectron} and \eqref{W:solantineutrino} in the general formula for the amplitude \eqref{AmpDecayW}:
\begin{align}
\mathcal{A}_{[{W^{-} \rightarrow e^{-} + \tilde{\nu}}]}^{[\lambda = \pm 1]} &= \frac{i g}{2\sqrt{2}} \int d^4x \sqrt{-g(x)}\bar{u}_{\vec{p}\sigma}(x) \gamma^{\hat{\alpha}} e^{i}_{\hat{j}}(1-\gamma^5) v_{\vec{p' \sigma'}}(x) f_{\vec{P}\lambda = \pm 1,_{i}}(x),
\end{align}
and then multiplying the terms.

The spatial integral is taken care of by:
\begin{equation}
\int d^3 x \,\, e^{-i\left(\vec{p}+\vec{p'}-\vec{P}\right)\vec x} = (2\pi)^3 \delta^{3}\left(\vec{p}+\vec{p'}-\vec{P}\right).
\end{equation}

What remains is a temporal integral of the form:
\begin{align}\label{AmpDecayWz}
\mathcal{A}_{[{W^{-} \rightarrow e^{-} + \tilde{\nu}}]}^{[\lambda=\pm 1]} &= -\frac{g}{4\sqrt{2}}\frac{\pi \sqrt{p} e^{\pi k/2} e^{-\pi K/2}}{(2\pi)^{3/2}}\left(\frac{1}{2}+\sigma' \right)\delta^3(\vec{p}+\vec{p'}-\vec{P}) \nonumber\\
&\times \int z\, dz\, e^{-ip' z}H_{\frac{1}{2}+ik}^{(2)}(pz)H_{iK}^{(1)}(Pz) \xi_{\sigma}^{\dag}(\vec{p})\sigma^{j}\eta_{\sigma'}(\vec{p'})\epsilon_{j}(\vec{P},\lambda=\pm 1),
\end{align}
where $z= -t_c = e^{-\omega t}/\omega$.

To solve the $z$ integral we will rewrite the exponential and Hankel function corresponding to the electron in terms of Bessel J functions, and the Hankel function corresponding to the $W^{-}$ boson in terms of the modified Bessel K function, using formulas \eqref{exptok} - \eqref{hankeltoj} from \hyperref[AppendixA]{Appendix A} \cite{AS,21}.

These transformations lead to:
\begin{align}\label{ampWdecaylambda1}
\mathcal{A}_{[{W^{-} \rightarrow e^{-} + \tilde{\nu}}]}^{[\lambda=\pm 1]} &= \frac{g\sqrt{i}}{4}\frac{\sqrt{\pi}\sqrt{p p'}e^{\pi k/2}e^{-i \pi/4}}{(2\pi)^{3/2} \,i \cosh{(\pi k)}} \left(\frac{1}{2}+\sigma'\right)\delta^3(\vec{p}+\vec{p'}-\vec{P})\nonumber\\
&\times \left[T_1 - T_2 - T_3 + T_4\right]\xi_{\sigma}^{\dag}(\vec{p})\sigma^{j}\eta_{\sigma'}(\vec{p'})\epsilon_{j}(\vec{P},\lambda=\pm 1),
\end{align}
where we now have the following integrals with respect to $z$:
\begin{align}\label{intJJK}
T_1 =& \int dz\, z\sqrt{z}\, i e^{-\pi k } J_{\frac{1}{2}}(p'z) J_{\frac{1}{2} +ik}(pz)  K_{iK}(-iPz), \nonumber\\
T_2 =& \int dz\, z\sqrt{z}\, J_{\frac{1}{2}}(p'z) J_{-\frac{1}{2} -ik}(pz)  K_{iK}(-iPz), \nonumber\\
T_3 =& \int dz\, z\sqrt{z}\, e^{-\pi k } J_{-\frac{1}{2}}(p'z) J_{\frac{1}{2} +ik}(pz)  K_{iK}(-iPz), \nonumber\\
T_4 =& \int dz\, z\sqrt{z}\, i^{-1} J_{-\frac{1}{2}}(p'z) J_{-\frac{1}{2} -ik}(pz)  K_{iK}(-iPz).
\end{align}

Integrals \eqref{intJJK} can be solved by using fromula \eqref{intBesseltoF4} from \hyperref[AppendixA]{Appendix A} \cite{AS,21}: 
\begin{align}
T_1 &= \frac{i e^{-\pi k}\sqrt{2} p'^{\frac{1}{2}} p^{\frac{1}{2}+ik} (-iP)^{-\frac{7}{2}-ik}}{\Gamma\left(\frac{3}{2}\right)\Gamma\left(\frac{3}{2}+ik\right)}  \Gamma\left(\frac{\frac{7}{2}+i(k-K)}{2}\right)  \Gamma\left(\frac{\frac{7}{2}+i(k+K)}{2}\right)\nonumber \\
&\times F_{4}\Bigg( \frac{\frac{7}{2}+i(k-K)}{2}, \frac{\frac{7}{2}+i(k+K)}{2}, \frac{3}{2}, \frac{3}{2} +ik,\frac{p'^2}{P^2},\frac{p^2}{P^2}\Bigg),\\
T_2 &= \frac{\sqrt{2} p'^{\frac{1}{2}} p^{-\frac{1}{2}-ik} (-iP)^{-\frac{5}{2}+ik}}{\Gamma\left(\frac{3}{2}\right)\Gamma\left(\frac{1}{2}-ik\right)}  \Gamma\left(\frac{\frac{5}{2}-i(k+K)}{2}\right)  \Gamma\left(\frac{\frac{5}{2}-i(k-K)}{2}\right)\nonumber \\
&\times F_{4}\Bigg(\frac{\frac{5}{2}-i(k+K)}{2}, \frac{\frac{5}{2}-i(k-K)}{2},\frac{3}{2}, \frac{1}{2} - ik,\frac{p'^2}{P^2},\frac{p^2}{P^2}\Bigg),\\
T_3 &= \frac{e^{-\pi k}\sqrt{2} p'^{-\frac{1}{2}} p^{\frac{1}{2}+ik} (-iP)^{-\frac{5}{2}-ik}}{\Gamma\left(\frac{1}{2}\right)\Gamma\left(\frac{3}{2}+ik\right)}  \Gamma\left(\frac{\frac{5}{2}+i(k-K)}{2}\right)  \Gamma\left(\frac{\frac{5}{2}+i(k+K)}{2}\right)\nonumber \\
&\times F_{4}\Bigg(\frac{\frac{5}{2}+i(k-K)}{2}, \frac{\frac{5}{2}+i(k+K)}{2},\frac{1}{2}, \frac{3}{2} + ik, \frac{p'^2}{P^2},\frac{p^2}{P^2}\Bigg),\\
T_4 &= \frac{i^{-1} \sqrt{2} p'^{-\frac{1}{2}} p^{-\frac{1}{2}-ik} (-iP)^{-\frac{3}{2}+ik}}{\Gamma\left(\frac{1}{2}\right)\Gamma\left(\frac{1}{2}-ik\right)} \Gamma\left(\frac{\frac{3}{2}-i(k+K)}{2}\right)  \Gamma\left(\frac{\frac{3}{2}-i(k-K)}{2}\right)\nonumber \\
&\times F_{4}\Bigg(\frac{\frac{3}{2}-i(k+K)}{2}, \frac{\frac{3}{2}-i(k-K)}{2},\frac{1}{2}, \frac{1}{2} - ik,\frac{p'^2}{P^2},\frac{p^2}{P^2}\Bigg),
\end{align}
where $F_4$ is the Appell hypergeometric function. The Appell hypergeometrc function can be defined as a double series, as can be seen in formula \eqref{F4series} in \hyperref[AppendixA]{Appendix A.}  \cite{AS,21}.

\subsection{\bf{The amplitude for $\lambda = 0$}}

To obtain the amplitude for $\lambda = 0$ we simply replace the Proca solutions \cite{2} \eqref{W:solProcaLambda0t} and \eqref{W:solProcaLambda0s} in amplitude \eqref{AmpDecayW}: 
\begin{align}
\mathcal{A}_{[{W^{-} \rightarrow e^{-} + \tilde{\nu}}]}^{[\lambda = 0]} &= \frac{i g}{2\sqrt{2}} \int d^4x \sqrt{-g(x)}\bar{u}_{\vec{p}\sigma}(x) \gamma^{\hat{\alpha}} e^{0}_{\hat{\alpha}}(1-\gamma^5) v_{\vec{p' \sigma'}}(x) f_{\vec{P}\lambda = 0,_{0}}(x) \nonumber\\
&+ \frac{i g}{2\sqrt{2}} \int d^4x \sqrt{-g(x)} \bar{u}_{\vec{p}\sigma}(x) \gamma^{\hat{\alpha}} e^{i}_{\hat{\alpha}}(1-\gamma^5) v_{\vec{p' \sigma'}}(x) f_{\vec{P}\lambda = 0,_{i}}(x). \nonumber\\
\end{align}

Since now we have both temporal and spatial components for the Proca solution \cite{2}, and in turn the spatial Proca solution \eqref{W:solProcaLambda0s} has two terms, we can split the amplitude into three terms, for convenience:
\begin{align}\label{a123}
\mathcal{A}_{[{W^{-} \rightarrow e^{-} + \tilde{\nu}}]}^{[\lambda = 0]} &= \,\,\,\mathcal{A}^{[\lambda=0](1)}_{[{W^{-} \rightarrow e^{-} + \tilde{\nu}}]} + \mathcal{A}^{[\lambda=0](2)}_{[{W^{-} \rightarrow e^{-} + \tilde{\nu}}]} + \mathcal{A}^{[\lambda=0](3)}_{[{W^{-} \rightarrow e^{-} + \tilde{\nu}}]},
\end{align}
where $A^{(1)}$ accounts for the temporal Proca solution \eqref{W:solProcaLambda0t}, while $A^{(2)}$ and $A^{(3)}$ account for both parts of the spatial Proca solution \eqref{W:solProcaLambda0s}.

First we solve for the temporal component using the same steps as in the previous section:
\begin{align}
\mathcal{A}^{[\lambda=0](1)}_{[{W^{-} \rightarrow e^{-} + \tilde{\nu}}]} &= -\frac{i g}{4\sqrt{2}}\frac{P\sqrt{p} e^{\pi k /2}e^{-\pi K/2}\omega}{\left(2\pi\right)^{3/2}M_W}\left(\frac{1}{2} + \sigma'\right) \delta^3\left(\vec{p} + \vec{p'} - \vec{P}\right)  \nonumber \\
&\times\int z^2 \, dz\, e^{-i p' z} H_{\nu_{+}}^{(2)}(pz)H_{iK}^{(1)}(Pz)\xi^{\dag}_{\sigma}(\vec{p}) \eta_{\sigma'}(\vec{p'}).
\end{align}

In terms of Bessel functions we arrive at:
\begin{align}\label{ampWlambda01}
\mathcal{A}^{[\lambda=0](1)}_{[{W^{-} \rightarrow e^{-} + \tilde{\nu}}]} &= \frac{g\sqrt{i}}{4}\frac{\omega}{M_W}\frac{P\sqrt{\pi}\sqrt{p p'}e^{\pi k/2}e^{-i \pi/4}}{(2\pi)^{3/2}\cosh{(\pi k)}} \left(\frac{1}{2}+\sigma'\right)\delta^3(\vec{p}+\vec{p'}-\vec{P})\nonumber\\
&\times \left[T_5 - T_6 - T_7 + T_8\right]\xi_{\sigma}^{\dag}(\vec{p})\eta_{\sigma'}(\vec{p'}),
\end{align}
where we have the following integrals:
\begin{align}
T_5 =& \int dz\, z^2\sqrt{z}\, i e^{-\pi k } J_{\frac{1}{2}}(p'z) J_{\frac{1}{2} +ik}(pz)  K_{iK}(-iPz), \nonumber\\
T_6 =& \int dz\, z^2\sqrt{z}\, J_{\frac{1}{2}}(p'z) J_{-\frac{1}{2} -ik}(pz)  K_{iK}(-iPz), \nonumber\\
T_7 =& \int dz\, z^2\sqrt{z}\, e^{-\pi k } J_{-\frac{1}{2}}(p'z) J_{\frac{1}{2} +ik}(pz)  K_{iK}(-iPz), \nonumber\\
T_8 =& \int dz\, z^2\sqrt{z}\, i^{-1} J_{-\frac{1}{2}}(p'z) J_{-\frac{1}{2} -ik}(pz)  K_{iK}(-iPz).
\end{align}

By applying formula \eqref{intBesseltoF4} to the above terms we obtain:
\begin{align}
T_5 &= \frac{i e^{-\pi k} 2^{3/2} p'^{\frac{1}{2}} p^{\frac{1}{2}+ik} (-iP)^{-\frac{9}{2}-ik}}{\Gamma\left(\frac{3}{2}\right)\Gamma\left(\frac{3}{2}+ik\right)} \Gamma\left(\frac{\frac{9}{2}+i(k-K)}{2}\right)  \Gamma\left(\frac{\frac{9}{2}+i(k+K)}{2}\right)\nonumber \\
&\times F_{4}\Bigg( \frac{\frac{9}{2}+i(k-K)}{2}, \frac{\frac{9}{2}+i(k+K)}{2}, \frac{3}{2}, \frac{3}{2} +ik,\frac{p'^2}{P^2},\frac{p^2}{P^2}\Bigg),\\
T_6 &= \frac{2^{3/2} p'^{\frac{1}{2}} p^{-\frac{1}{2}-ik} (-iP)^{-\frac{7}{2}+ik}}{\Gamma\left(\frac{3}{2}\right)\Gamma\left(\frac{1}{2}-ik\right)} \Gamma\left(\frac{\frac{7}{2}-i(k+K)}{2}\right)  \Gamma\left(\frac{\frac{7}{2}-i(k-K)}{2}\right)\nonumber \\
&\times F_{4}\Bigg(\frac{\frac{7}{2}-i(k+K)}{2}, \frac{\frac{7}{2}-i(k-K)}{2}, \frac{3}{2}, \frac{1}{2} - i k,\frac{p'^2}{P^2},\frac{p^2}{P^2}\Bigg),\\
T_7 &= \frac{e^{-\pi k}2^{3/2} p'^{-\frac{1}{2}} p^{\frac{1}{2}+ik} (-iP)^{-\frac{7}{2}-ik}}{\Gamma\left(\frac{1}{2}\right)\Gamma\left(\frac{3}{2}+ik\right)}  \Gamma\left(\frac{\frac{7}{2}+i(k-K)}{2}\right)  \Gamma\left(\frac{\frac{7}{2}+i(k+K)}{2}\right)\nonumber \\
&\times F_{4}\Bigg(\frac{\frac{7}{2}+i(k-K)}{2}, \frac{\frac{7}{2}+i(k+K)}{2}, \frac{1}{2}, \frac{3}{2} + ik,\frac{p'^2}{P^2},\frac{p^2}{P^2}\Bigg),\\
T_8 &= \frac{i^{-1} 2^{3/2} p'^{-\frac{1}{2}} p^{-\frac{1}{2}-ik} (-iP)^{-\frac{5}{2}+ik}}{\Gamma\left(\frac{1}{2}\right)\Gamma\left(\frac{1}{2}-ik\right)}  \Gamma\left(\frac{\frac{5}{2}-i(k+K)}{2}\right)  \Gamma\left(\frac{\frac{5}{2}-i(k-K)}{2}\right)\nonumber \\
&\times F_{4}\Bigg(\frac{\frac{5}{2}-i(k+K)}{2}, \frac{\frac{5}{2}-i(k-K)}{2},\frac{1}{2}, \frac{1}{2} - ik,\frac{p'^2}{P^2},\frac{p^2}{P^2}\Bigg).
\end{align}

The second term in the amplitude \eqref{a123} ends up being the amplitude for $\lambda=\pm 1$ up to an imaginary factor and ratio $\omega/M_W$, but with a different polarization vector and different helicity spinors: 
\begin{align}\label{ampWlambda02}
\mathcal{A}^{[\lambda=0](2)}_{[{W^{-} \rightarrow e^{-} + \tilde{\nu}}]} = - \frac{i \,\omega}{M_W} \left(\frac{1}{2} +iK\right)  \mathcal{A}(\pm1)   \xi_{\sigma}^{\dag}(\vec{p})\sigma^{j}\eta_{\sigma'}(\vec{p'})\epsilon_{j}(\vec{P},\lambda=0),
\end{align}
where we denote:
\begin{align}
\mathcal{A}(\pm 1) = \frac{g\sqrt{i}}{4}\frac{\sqrt{\pi}\sqrt{p p'}e^{\pi k/2}e^{-i \pi/4}}{(2\pi)^{3/2} \,i \cosh{(\pi k)}} \left(\frac{1}{2}+\sigma'\right)\delta^3(\vec{p}+\vec{p'}-\vec{P})\left[T_1 - T_2 - T_3 + T_4\right],
\end{align}
since we can rewrite the amplitude for $\lambda=\pm1$ \eqref{ampWdecaylambda1} as such:
\begin{align}\label{ampWpm1}
\mathcal{A}_{[{W^{-} \rightarrow e^{-} + \tilde{\nu}}]}^{[\lambda=\pm 1]} &= \mathcal{A}(\pm1) \xi_{\sigma}^{\dag}(\vec{p})\sigma^{j}\eta_{\sigma'}(\vec{p'})\epsilon_{j}(\vec{P},\lambda=\pm 1).
\end{align}

We are left with the third term:
\begin{align}\label{AmpDecayLambda1Hankel}
\mathcal{A}^{[\lambda=0](3)}_{[{W^{-} \rightarrow e^{-} + \tilde{\nu}}]} &= -\frac{i g}{4\sqrt{2}}\frac{P\sqrt{p}  e^{\pi k /2}e^{-\pi K/2}\omega}{\left(2\pi\right)^{3/2}M_W} \left(\frac{1}{2} + \sigma'\right) \delta^3\left(\vec{p} + \vec{p'} - \vec{P}\right)  \nonumber \\
&\times\int z^2 \, dz\, e^{-i p' z} H_{\nu_{+}}^{(2)}(pz)H_{1+iK}^{(1)}(Pz)\xi_{\sigma}^{\dag}(\vec{p})\sigma^{j}\eta_{\sigma'}(\vec{p'})\epsilon_{j}(\vec{P},\lambda=\pm 1).
\end{align}

After transforming the exponential and the Hankel funtions from \eqref{AmpDecayLambda1Hankel} into Bessel J and $K$ functions \cite{AS,21}, the amplitude can be written as:
\begin{align}\label{ampWlambda03}
\mathcal{A}^{[\lambda=0](3)}_{[{W^{-} \rightarrow e^{-} + \tilde{\nu}}]} &= -\frac{i \sqrt{i}g}{4}\frac{P\sqrt{\pi}\sqrt{p p'}  e^{-i\pi/4} e^{\pi k /2} \omega}{\left(2\pi\right)^{3/2}M_W \cosh{(\pi k)}}\left(\frac{1}{2} + \sigma'\right) \delta^3\left(\vec{p} + \vec{p'} - \vec{P}\right)  \nonumber \\
&\times\left[T_9 -T_{10}-T_{11}+T_{12} \right] \xi_{\sigma}^{\dag}(\vec{p})\sigma^{j}\eta_{\sigma'}(\vec{p'})\epsilon_{j}(\vec{P},\lambda=\pm 1),
\end{align}
where terms $T_9 - T_{12}$ are defined below:
\begin{align}
T_9 =& \int dz\, z^2\sqrt{z}\, i e^{-\pi k } J_{\frac{1}{2}}(p'z) J_{\frac{1}{2} +ik}(pz)  K_{1+iK}(-iPz), \nonumber\\
T_{10} =& \int dz\, z^2\sqrt{z}\, J_{\frac{1}{2}}(p'z) J_{-\frac{1}{2} -ik}(pz)  K_{1+iK}(-iPz), \nonumber\\
T_{11} =& \int dz\, z^2\sqrt{z}\, e^{-\pi k } J_{-\frac{1}{2}}(p'z) J_{\frac{1}{2} +ik}(pz)  K_{1+iK}(-iPz), \nonumber\\
T_{12} =& \int dz\, z^2\sqrt{z}\, i^{-1} J_{-\frac{1}{2}}(p'z) J_{-\frac{1}{2} -ik}(pz)  K_{1+iK}(-iPz).
\end{align}

After solving the integrals the terms become:
\begin{align}
T_9 &= \frac{i e^{-\pi k} 2^{3/2} p'^{\frac{1}{2}} p^{\frac{1}{2}+ik} (-iP)^{-\frac{9}{2}-ik}}{\Gamma\left(\frac{3}{2}\right)\Gamma\left(\frac{3}{2}+ik\right)}  \Gamma\left(\frac{\frac{7}{2}+i(k-K)}{2}\right)  \Gamma\left(\frac{\frac{11}{2}+i(k+K)}{2}\right)\nonumber \\
&\times F_{4}\Bigg( \frac{\frac{7}{2}+i(k-K)}{2}, \frac{\frac{11}{2}+i(k+K)}{2}, \frac{3}{2}, \frac{3}{2} +ik,\frac{p'^2}{P^2},\frac{p^2}{P^2}\Bigg),\\
T_{10} &= \frac{2^{3/2} p'^{\frac{1}{2}} p^{-\frac{1}{2}-ik} (-iP)^{-\frac{7}{2}+ik}}{\Gamma\left(\frac{3}{2}\right)\Gamma\left(\frac{1}{2}-ik\right)} \Gamma\left(\frac{\frac{5}{2}-i(k+K)}{2}\right)  \Gamma\left(\frac{\frac{9}{2}-i(k-K)}{2}\right)\nonumber \\
&\times F_{4}\Bigg(\frac{\frac{5}{2}-i(k+K)}{2}, \frac{\frac{9}{2}-i(k-K)}{2}, \frac{3}{2}, \frac{1}{2} - i k,\frac{p'^2}{P^2},\frac{p^2}{P^2}\Bigg),\\
T_{11} &= \frac{e^{-\pi k}2^{3/2} p'^{-\frac{1}{2}} p^{\frac{1}{2}+ik} (-iP)^{-\frac{7}{2}-ik}}{\Gamma\left(\frac{1}{2}\right)\Gamma\left(\frac{3}{2}+ik\right)} \Gamma\left(\frac{\frac{5}{2}+i(k-K)}{2}\right)  \Gamma\left(\frac{\frac{9}{2}+i(k+K)}{2}\right)\nonumber \\
&\times F_{4}\Bigg(\frac{\frac{5}{2}+i(k-K)}{2}, \frac{\frac{9}{2}+i(k+K)}{2}, \frac{1}{2}, \frac{3}{2} + ik,\frac{p'^2}{P^2},\frac{p^2}{P^2}\Bigg),\\
T_{12} &= \frac{i^{-1} 2^{3/2} p'^{-\frac{1}{2}} p^{-\frac{1}{2}-ik} (-iP)^{-\frac{5}{2}+ik}}{\Gamma\left(\frac{1}{2}\right)\Gamma\left(\frac{1}{2}-ik\right)}\Gamma\left(\frac{\frac{3}{2}-i(k+K)}{2}\right)  \Gamma\left(\frac{\frac{7}{2}-i(k-K)}{2}\right) \nonumber\\
&\times F_{4}\Bigg(\frac{\frac{3}{2}-i(k+K)}{2}, \frac{\frac{7}{2}-i(k-K)}{2}, \frac{1}{2}, \frac{1}{2} - ik,\frac{p'^2}{P^2},\frac{p^2}{P^2}\Bigg).
\end{align}

It is worth pointing out that all three terms \eqref{ampWlambda01},\eqref{ampWlambda02}  and \eqref{ampWlambda03} that make up the amplitude for $\lambda = 0$ have a factor of $\omega/M_W$. This means that the only term which doesn't vanish when the Hubble parameter $\omega \rightarrow 0$ and there is no spacetime expansion is \eqref{ampWlambda02}. Therefore, this will be the only term with longitudinal polarisation which survives in the Minkowski limit, as we will see in section \ref{sec2.4}.

\subsection{\bf{Total probability and transition rate in the large expansion limit}}

In the early Universe the $W$ bosons were generated from vacuum and also produced in emission processes by fermions, but they began to decay when the background energy became smaller than their rest energy. These phenomena happened in a very short time. One of the important applications of our general computations in de Sitter spacetime is to obtain the transition rates of the processes we study. This can be done in the limit of a large expansion factor $\omega \gg m_e,\, \omega \gg M_W$, a limit which can be identified with the very early universe. Thus we can take the ratios between the particle masses and Hubble parameter to be zero throughout all our equations for the amplitude, that is: $m_e/\omega \rightarrow 0, M_W/\omega \rightarrow 0$. 

This brings a simplification to the amplitude formula \eqref{AmpDecayWz}, in the case of transversal polarisation $\lambda = \pm 1$:
\begin{align}\label{AmpWrata1}
\mathcal{A}_{[W^{-}\rightarrow {e^{-} +\widetilde{\nu} }]}^{[\lambda=\pm1]} &= \frac{-g}{4\sqrt{2}}\frac{\pi\sqrt{p}e^{-i\pi/4}e^{-i\pi/4}}{(2\pi)^{3/2}}  \left(\frac{1}{2}+\sigma'\right) \delta^3(\vec{P}-\vec{p}-\vec{p}\,'\,) \nonumber \\
&\times \int_{0}^{\infty} dz\, z e^{-i p\, 'z}H_{1/2}^{(2)} (pz)H_{1/2}^{(1)}(Pz) \xi^{\dag}_{\sigma}(\vec{p})\vec{\sigma}\cdot\vec{\epsilon}(\vec{n}_{\mathcal{P}},\lambda)\eta_{\sigma'}(\vec{p}\,'),
\end{align}
where the order of the Hankel function corresponding to the electron of momentum $p$ in \eqref{AmpDecayWz} has become $1/2$, since $k=m_e/\omega = 0$. The order of the Hankel function corresponding to the $W^{-}$ of momentum $P$ boson also goes to 1/2:
\begin{equation}
iK = i\sqrt{\left(\frac{M_W}{\omega}\right)^{2}-\frac{1}{4}} = \sqrt{\frac{1}{4} - \left(\frac{M_W}{\omega}\right)^{2}} \rightarrow \frac{1}{2}, \nonumber
\end{equation}
in the limit when $M_W/\omega \rightarrow 0$. Since the ampltiude for logitudinal polarisation $\lambda = 0$ has an overall factor of $\omega/M_W$, as seen in the previous section, we can see that it diverges in this large expansion limit, and in consequence does not contribute to the transition rate we are computing here.

The amplitude can be split into factors as such:
\begin{align}
\mathcal{A}_{[W^{-}\rightarrow {e^{-} +\widetilde{\nu} }]}^{[\lambda=\pm1]} =\mathcal{I}_{if}
\mathcal{M}_{if}\delta^3(\vec{P}-\vec{p}-\vec{p}\,'\,),
\end{align}
where $\mathcal{I}_{if}$ denotes the temporal integral:
\begin{equation}\label{IifrataW}
\mathcal{I}_{if} = \int_{0}^{\infty} dz\, z\, e^{-i p\, 'z}H_{1/2}^{(2)} (pz)H_{1/2}^{(1)}(Pz), 
\end{equation}
while $\mathcal{M}_{if}$ contains the rest of the factors in \eqref{AmpWrata1}. 

The temporal integral can be solved by transforming the Hankel functions into exponentials by using first formula \eqref{hankeltok} from \hyperref[Appendix A]{Appendix A}, and then formula \eqref{exptok} in reverse, which leads to:
\begin{align}
\mathcal{I}_{if} = \frac{2}{\pi \sqrt{pP}} \int_{0}^{\infty} dz \, e^{i(P-p'-p)z}.
\end{align}

We can integrate over $z$ afterwards:
\begin{align}
\mathcal{I}_{if} = \frac{-2i}{\pi\sqrt{pP}(p + p' - P)}.
\end{align}

Following this procedure the amplitude reads:
\begin{align}\label{AmpWlargeexp}
\mathcal{A}_{[{W^{-}\rightarrow e^{-} +\widetilde{\nu}}]}^{[\lambda=\pm 1]} &= \frac{i g}{2\sqrt{2}}\frac{e^{-i\pi/2}}{(2\pi)^{3/2}} \frac{\delta^3(\vec{P}-\vec{p}-\vec{p}\,'\,) }{\sqrt{P}(p+p'-P)} \xi^{\dag}_{\sigma}(\vec{p})\vec{\sigma}\cdot\vec{\epsilon}(\vec{n}_{\mathcal{P}},\lambda)\eta_{\sigma'}(\vec{p}\,'),
\end{align}
where we have taken into account that the helicity of the antineutrino has to be $\sigma'=1/2$, otherwise \eqref{AmpWrata1} would cancel out.

The transition rate has been defined in previous studies related to the large expansion limit in this de Sitter setup, and a complete discussion about this subject can be found in \cite{cpc}.  Here we give just the definition, which has been adapted to our case:
\begin{align}\label{ratadefW}
R_{i\rightarrow f} & = \frac{\delta^3 (\vec{P} - \vec{p} - \vec{p}\,') }{2(2\pi)^3} \sum_{\sigma} |\mathcal{M}_{if} |^2 |\mathcal{I}_{if}|  \lim_{t\rightarrow \infty} |e^{\omega t} \mathcal{K}_{i\rightarrow f}| , 
\end{align}
where $\mathcal{K}_{i\rightarrow f}$ is the integrand of the temporal integral:
\begin{equation}
\mathcal{K}_{i\rightarrow f} = z\, e^{-i p'z} H_{1/2}^{(2)}(pz) H_{1/2}^{(1)}(Pz),
\end{equation}
while:
\begin{equation}
\mathcal{M}_{if} = \frac{i g}{2\sqrt{2}}\frac{\pi\sqrt{p}e^{-i\pi/2}}{(2\pi)^{3/2}} \xi^{\dag}_{\sigma}(\vec{p})\vec{\sigma}\cdot\vec{\epsilon}(\vec{n}_{\mathcal{P}},\lambda)\eta_{1/2}(\vec{p'}).
\end{equation}

Next we solve the temporal limit in \eqref{ratadefW}, keeping in mind that $z=-t_c=e^{-\omega t}/\omega$:
\begin{align}
\lim_{t\rightarrow t_{\infty}} |e^{\omega t} \mathcal{K}_{i\rightarrow f}|&= \lim_{t\rightarrow t_{\infty}} \frac{2}{\pi \sqrt{Pp}}\bigg| e^{\omega t} e^{-\omega t} e^{i(P-p-p')\exp{(-\omega t})/\omega}\bigg| \nonumber\\
&=\frac{2}{\pi\sqrt{Pp}},
\end{align}
where by $t_{\infty}$ we denote a sufficiently large time after the interaction.

The transition rate becomes:
\begin{align}
R_{_{i\rightarrow f}}&=\frac{g^2}{8(2\pi)^6}  \frac{\delta^3(\vec{P}-\vec{p}-\vec{p}\,'\,) }{P(p+p'-P)} \sum_{\sigma\lambda}\big|\xi^{\dag}_{\sigma}(\vec{p})\vec{\sigma}\cdot\vec{\epsilon}(\vec P,\lambda)\eta_{1/2}(\vec{p}\,')\big|^2.
\end{align}

The summation of the helicity spinors and polarization vectors is straightforward: we set the particles on the same direction, i.e. the third axis such that $\vec P=P \vec e_3;\,\vec p=p \vec e_3;\, \vec{p'} = -p' \vec e_3$. The bispinor summation has just one nonvanishing term, if one uses circular polarizations:
\begin{equation}
\xi^{\dag}_{-1/2}(\vec{p})\vec{\sigma}\cdot\vec{\epsilon}_{-1}(\vec P,\lambda)\eta_{1/2}(\vec{p}\,')=-\sqrt{2}. 
\end{equation}

In consequence, the final form of the rate is:
\begin{equation}
R_{_{i\rightarrow f}}=\frac{g^2}{4(2\pi)^6}  \frac{\delta^3(\vec{P}-\vec{p}-\vec{p}\,'\,) }{P(p+p'-P)}  .
\end{equation}

The total transition rate for the $W^{-}$ decay in this limit is obtained by integrating after the final momenta of the electron and anti-neutrino:
\begin{align}\label{rataWtot}
R_{W^-\rightarrow e^-+\widetilde{\nu}} &= \int d^3p \int d^3 p' \, R_{i\rightarrow f} \nonumber\\
&= \int d^3p \int d^3 p' \frac{g^2}{4(2\pi)^6}  \frac{\delta^3(\vec{P}-\vec{p}-\vec{p}\,'\,) }{P(p+p'-P)}.
\end{align}

The momenta integrals are divergent and from the known regularization methods \cite{WF,GHV,IT,GT,PV,HGF}, it seems that dimensional regularization \cite{GHV,IT,HGF} will work best moving forward. The first step is to solve the momentum integral with respect to $p'$:
\begin{align}
I &= \frac{1}{(2\pi)^3} \int d^3p \int d^3p' \frac{\delta^3 (\vec{P} - \vec{p} - \vec{p}\,')}{P(p+p'-P)} \nonumber\\
&= \frac{1}{(2\pi)^3} \int d^3p \frac{1}{2P(p-P)}
\end{align}

The above integral is divergent, so we can write it in $D$ dimensions:
\begin{align}\label{id4}
I(D) = \frac{2\pi^{D/2}}{2P (2\pi)^D \Gamma(D/2)} \int_{0}^{\infty} dp \frac{p^{D-1}}{p-P}.
\end{align}

Next we change the integration variable to $p=-Py$, which leads to:
\begin{align}\label{id5}
I(D) = \frac{2\pi^{D/2}(-P)^{D-1}}{2P(2\pi)^D \Gamma(D/2)} \int_{0}^{\infty} \frac{y^{D-1}}{1+y} dy. 
\end{align}

Integral \eqref{id5} can be solved by using the Beta Euler function \eqref{betaeuler}:
\begin{align}\label{IDdim}
I(D) = \frac{2\pi^{D/2}(-P)^{D-1} \Gamma(D) \Gamma(1-D)}{(2\pi)^D 2P \Gamma(D/2)},
\end{align}
for $\alpha=D$ and $\beta=1-D$.

However, this integral is still divergent in $\Gamma(1-D)$ for $D=3$. To remove this divergence one can apply the minimal subtraction method \cite{GT}, and compute the residue of the integral in $D=3$:
\begin{equation}
\text{Res}\, I(D)= \lim_{D\rightarrow 3} (3-D) I(D) = -\frac{P}{4\pi^2}.
\end{equation}

Afterwards we choose a counter-term of the form:
\begin{equation}
\centering   \frac{\mu^s \,\text{Res} \,I(D)}{3-D},
\end{equation}
where $\mu$ is a mass parameter and $s$ is taken such that this term has the same dimension as $I(D)$.

Then we can define the regularized integral as such:
\begin{equation}
I(D)_r = I(D) - \frac{P\mu^{D-3}}{4\pi^2 (3-D)}.
\end{equation}

Explicitly this is:
\begin{align}\label{IDrdimsub}
I(D)_r =  \frac{1}{3-D}\bigg( \frac{2\pi^{D/2}(-P)^{D-1} \Gamma(D) \Gamma(4-D)}{2P(2\pi)^D (1-D)(2-D) \Gamma(D/2)} 
-\frac{P\mu^{D-3}}{4\pi^2} \bigg),
\end{align}
where we have used $\Gamma(z+1) = z\Gamma(z)$ to transform $\Gamma(1-D)$ in $I(D)$ \eqref{IDdim}.

By expanding the paranthesis around the value $D=3$ (we use formula \eqref{gammaexpansion} for the expansion of the Gamma functions) we obtain:
\begin{align}\label{id5}
\frac{2\pi^{D/2}(-P)^{D-1} \Gamma(D) \Gamma(4-D)}{2P(2\pi)^D (1-D)(2-D) \Gamma(D/2)} - \frac{P\mu^{D-3}}{4\pi^2} = \frac{P(3-D)}{8\pi^2} \bigg[ \ln \Big( \frac{4\pi\mu^2}{P^2} \Big) + \psi \Big(\frac{3}{2}\Big) \bigg] + \mathcal{O}\big((D-3)^2\big).
\end{align}

Now the factor of $3-D$ in front of the parenthesis will simplify the divergent term $1/(3-D)$ in \eqref{IDrdimsub}. 

We obtain the final result for the transition rate after taking the limit $D=3$ in \eqref{IDrdimsub} and replacing the result in \eqref{rataWtot}, while taking into account the remaining factors:
\begin{align}\label{deca1}
R_{W^-\rightarrow e^-+\widetilde{\nu}} 
=\frac{ G_FM_W^2\, P}{4\sqrt{2}\pi^2(2\pi)^3 } \bigg[ \ln \Big( \frac{4\pi\mu^2}{P^2} \Big) + \psi \Big(\frac{3}{2}\Big) \bigg].
\end{align}

The Fermi constant $G_F$ is related to the coupling constant by relation:
\begin{equation}\label{ctcu}
g^2 = 4\sqrt{2} G_{F} M_{W}^2.
\end{equation}

The total transition rate in the limit of large expansion is finite and depends on the momentum $P$ and mass of the $W$ boson. 

The probability per unit volume can also be computed by taking the squared modulus of amplitude \eqref{AmpWlargeexp} and by using relation $\big|(2\pi)^3\delta^3(\vec p)\big|^2=(2\pi)^3V\delta^3(\vec p)$:
\begin{align} 
P_{i\rightarrow f} &= \frac{1}{2}\sum_{\sigma\lambda}\frac{g^2}{8(2\pi)^3}  \frac{\delta^3(\vec{P}-\vec{p}-\vec{p}\,'\,) }{P(p+p'-P)^2}\big|\xi^{\dag}_{\sigma}(\vec{p})\vec{\sigma}\cdot\vec{\epsilon}(\vec P,\lambda)\eta_{1/2}(\vec{p}\,')\big|^2.
\end{align}

The total probability of $W^{-}$ decay in the large expanstion limit can be obtained by integrating after the final momenta of the particles: 
\begin{equation}
P_{W^-\rightarrow e^-+\widetilde{\nu}} = \int d^3p \int d^3p'  P_{i\rightarrow f}.
\end{equation}

The momenta integrals that need to be solved are of the type:
\begin{align}
I &= \frac{1}{(2\pi)^3} \int d^3p \int d^3p' \frac{\delta^3 (\vec{P} - \vec{p} - \vec{p}\,')}{P(p+p'-P)^2} \nonumber\\
&= \frac{1}{(2\pi)^3} \int d^3p \frac{1}{4P(p-P)^2}.
\end{align}

By applying the dimensional regularization together with the minimal subtraction method \cite{WF,GHV,IT,GT,HGF}, as in the case of the total transition rate, the final result turns out to be:
\begin{equation}
P_{W^-\rightarrow e^-+\widetilde{\nu}} =   \frac{G_FM_W^2}{8\sqrt{2}\pi^2}\bigg[ 1-  \psi \Big(\frac{3}{2} \Big) - \ln \Big(\frac{4\pi \mu^2}{P^2} \Big) \bigg]. 
\end{equation}

From the above formula we can see that the total probability is well defined when $4 \pi \mu^2 \leq P^2$.

The results from this section prove that renormalization techniques used in flat spacetime field theory can be applied in a curved background. We have obtained a finite result for the decay rate of the $W^{-}$ boson in the de Sitter geometry, and we observe that this rate \eqref{deca1}, computed in the high expansion limit, differs from the one computed in Minkowski spacetime for the same process \eqref{decaminkw}. This difference will be discussed further in the next section.

\subsection{\bf{The Minkowski limit}}\label{sec2.4}

Since the decays of $W$ bosons into lepton pairs are well studied in Minkowski electro-weak theory \cite{12}, one would expect to recover the decay rates of these processes in flat spacetime from the decay rates in an expanding background, as a limiting case. However, the general result for the amplitudes is expressed in terms of Appell $F_4$ functions, which are not well studied in the literature. This poses challenges when trying to recover the Minkowski limit of the decay rates. 

There have been some results regarding integrals involving a power function and three Bessel functions of integer orders \cite{auluk}. Here the result of this type of integrals is obtained as a semi-empirical approximate formula in terms of the Dirac delta functions \cite{auluk}. However, in our computations we are dealing with Hankel functions of imaginary order. Consequently we can only obtain the Minkowski limit when neglecting the fermion masses, in order for the imaginary part of the Hankel function order to vanish. This corresponds physically to the ultrarelativistic limit, i.e. the momenta of the particles are much larger than their masses.

In trying to recover the Minkowski limit of the transition rate, we see that there will be two contributions to the amplitude, one from the amplitude for $\lambda = \pm1$ and one from the amplitude for $\lambda=0$:
\begin{equation}\label{AM}
\mathcal{A}_{[W^-\rightarrow e^- + \tilde{\nu}]}^{[\omega \rightarrow 0]} = \mathcal{A}_{[W^-\rightarrow e^- + \tilde{\nu}]}^{[\lambda=\pm1]} + \mathcal{A}^{[\lambda=0](2)}_{[{W^{-} \rightarrow e^{-} + \tilde{\nu}}]},
\end{equation}
where we have denoted $\mathcal{A}_{[W^-\rightarrow e^- + \tilde{\nu}]}^{[\omega \rightarrow 0]}$ the amplitude in the Minkowski limit.

Only the second term from the longitudinal polarisation amplitude \eqref{a123} survives in the Minkowski limit when $\omega \rightarrow 0$, since all three terms which make up the amplitude for $\lambda = 0$, \eqref{ampWlambda01}, \eqref{ampWlambda02}, \eqref{ampWlambda03}, are multiplied by a factor of $\omega/M_W$. 

For the second term, however, we have a more complicated dependence on $\omega/M_W$ through $K$:
\begin{align}\label{ampWlambda022}
\mathcal{A}^{[\lambda=0](2)}_{[{W^{-} \rightarrow e^{-} + \tilde{\nu}}]} = - \frac{i \,\omega}{M_W} \left(\frac{1}{2} +iK\right)  \mathcal{A}(\pm1)   \xi_{\sigma}^{\dag}(\vec{p})\sigma^{j}\eta_{\sigma'}(\vec{p'})\epsilon_{j}(\vec{P},\lambda=0),
\end{align}
where $K$ is defined as:
\begin{align}
K = \sqrt{\left(\frac{M_W}{\omega}\right)^2 - \frac{1}{4}}. 
\end{align}

Therefore we can write:
\begin{align}
- i \frac{\omega}{M_W}\left(\frac{1}{2} +iK\right) = -i \frac{\omega}{2 M_W} + \sqrt { 1 - \frac{1}{4} \frac{\omega^2}{M_W^2}}.
\end{align}

When $\omega \rightarrow 0$, term \eqref{ampWlambda022} becomes:
\begin{align}
\mathcal{A}^{[\lambda=0](2)}_{[{W^{-} \rightarrow e^{-} + \tilde{\nu}}]} = \mathcal{A}(\pm1)   \xi_{\sigma}^{\dag}(\vec{p})\sigma^{j}\eta_{\sigma'}(\vec{p'})\epsilon_{j}(\vec{P},\lambda=0).
\end{align}

By using the above formula, and also formula \eqref{ampWpm1}, we can rewrite the Minkowski amplitude:
\begin{align}
\mathcal{A}_{[W^-\rightarrow e^- + \tilde{\nu}]}^{[\omega \rightarrow 0]} = \mathcal{A}(\pm1) \left[\xi_{\sigma}^{\dag}(\vec{p})\sigma^{j}\eta_{\sigma'}(\vec{p'})\epsilon_{j}(\vec{P},\lambda= \pm1) + \xi_{\sigma}^{\dag}(\vec{p})\sigma^{j}\eta_{\sigma'},(\vec{p'})\epsilon_{j}(\vec{P},\lambda= 0)\right],
\end{align}
where
\begin{align}
\mathcal{A}(\pm 1) &=  -\frac{g}{4\sqrt{2}}\frac{\pi \sqrt{p} e^{\pi k/2} e^{-\pi K/2}}{(2\pi)^{3/2}}\left(\frac{1}{2}+\sigma' \right)\delta^3(\vec{p}+\vec{p'}-\vec{P}) \nonumber\\
&\times \int z\, dz\, e^{-ip' z}H_{\frac{1}{2}+ik}^{(2)}(pz)H_{iK}^{(1)}(Pz).
\end{align}

Taking the ultrarelativistic limit and neglecting the particle masses means $k=m_e/\omega \rightarrow 0$. Then amplitude \eqref{AmpDecayWz} can be rewritten by transforming the temporal integral according to formulas \eqref{limminkwt1}, \eqref{limminkwt2} and \eqref{kjj} from \hyperref[AppendixB]{Appendix B}:
\begin{align}\label{amm}
\mathcal{A}(\pm1) &= \frac{ig}{2\sqrt{2}}\left(\frac{1}{2}+\sigma'\right)\delta^3(\vec{P}-\vec{p}-\vec{p}\,'\,) \nonumber\\
&\times\sqrt{p+p'}\int_{0}^{\infty} dz \,z \, K_{iK} (-iPz)\bigg( J_{\frac{1}{2}}\big((p+p')z\big) + i J_{-\frac{1}{2}} \big((p+p')z\big) \bigg).
\end{align}

The details related to solving the temporal integral are given in \hyperref[Appendix B]{Appendix B} and we restrict ourselves to giving the final form of the amplitude in this limit. 

In the case when $p+p'=P$, the temporal integral reduces to:
\begin{align}
\lim_{\epsilon \rightarrow 0} I = \frac{i\pi e^{i\pi/4}}{2} \delta(p+p'-P)
\bigg( \frac{p+p'}{P\sqrt{P}}   + \frac{1}{\sqrt{P}}  \bigg).
\end{align}

With this result, the amplitude when $p+p'=P$ is:
\begin{align}\label{ampminkwdeltap}
\mathcal{A}(\pm1) &= \frac{ig\pi\,e^{i\pi/4}}{4\sqrt{2}(2\pi)^{3/2}}\,\delta^3(\vec{P}-\vec{p}-\vec{p}\,'\,)\left(\frac{1}{2}+\sigma'\right) \xi^{\dag}_{\sigma}(\vec{p})\vec{\sigma}\cdot\vec{\epsilon}(\vec{n}_{\mathcal{P}},\lambda)\eta_{\sigma'}(\vec{p}\,')\nonumber\\
&\times\frac{\delta(P-p-p')}{\sqrt{|\vec P|}}\left(\frac{|\vec p\,|+|\vec p\,'|}{|\vec P|}+1\right).
\end{align}

The complete result of the temporal integral from eq. \eqref{amm} is given in \hyperref[AppendixB]{Appendix B} and has, besides the delta Dirac term for $p+p'=P$, other terms proportional with the unit step functions for the case when $p+p'\neq P$. This result was obtained by using the principal part prescription. Moving forward we will use only the part of the amplitude proportional with the delta Dirac function and neglect the terms with $p+p'\neq P$, since they vanish in the Minkowski limit. 

It is well known that, in the Minkowski field theory, energy and momentum are conserved simultaneously \cite{12,19,20}. The Minkowski amplitudes \cite{12,19,20}, depend on the four dimensional delta function $\delta^4(p)=\delta (E)\delta^3(\vec P)$. In our case the amplitude is proportional only with the delta function that contains the momenta \eqref{ampminkwdeltap}. In this situation we will take the ultra-relativistic limit, when the energy is approximately equal with the modulus of the momentum for the massive particles. Thus the Dirac delta function which depends on the momenta can be turned into a Dirac delta function of energy.

The de Sitter amplitude in the ultra-relativistic case then reads as:
\begin{align}\label{amin}
\mathcal{A}_{[W^-\rightarrow e^- + \tilde{\nu}]}^{[\omega \rightarrow 0]} &= \frac{ig\,e^{i\pi/4}(2\pi)}{8 \sqrt{2}(2\pi)^{3/2}}\,\delta^3(\vec{P}-\vec{p}-\vec{p}\,'\,)\left(\frac{1}{2}+\sigma'\right) \frac{\delta(\mathcal{E}-E-E')}{\sqrt{|\vec P|}}\left(\frac{|\vec p\,|+|\vec p\,'|}{|\vec P|}+1\right) \nonumber\\
&\times \left[\xi^{\dag}_{\sigma}(\vec{p})\vec{\sigma}\cdot\vec{\epsilon}(\vec{n}_{\mathcal{P}},\lambda=\pm1)\eta_{\sigma'}(\vec{p}\,') + \xi^{\dag}_{\sigma}(\vec{p})\vec{\sigma}\cdot\vec{\epsilon}(\vec{n}_{\mathcal{P}},\lambda=0)\eta_{\sigma'}(\vec{p}\,') \right]
\end{align}
where $E,E'$ are the energies of electron and neutrino, and $\mathcal{E}=\sqrt{M_W^2+\vec P^2}$ is the energy of the $W$ boson in Minkowski space \cite{12,19,20}. We denote through this section the momenta modulus by $|\vec p|,|\vec p\,'|, |\vec P|$, as to not confuse it with the notation for the four momentum in Minkowski spacetime. The Minkowski limit is taken in the ultra-relativistic case where $\mathcal{E}=|\vec{P}|$, and the delta function of momenta transforms into a delta function of energy. This allows us to make the connection with the results from the Minkowski theory \cite{12,19,20}.

Amplitude \eqref{amin} needs to be summed over helicities $\sigma = \pm 1/2$ and polarisations $\lambda=-1,0,1$, while the helicity of the neutrino $\sigma' = 1/2$ remains fixed. We consider the case where the electron and antineutrino are emitted in the same direction, but their momenta have opposite orientations. In this case the momenta vectors have the components $\vec p(p,\pi,0),\,\vec p'(p',0,0)$ in polar coordinates. 

By taking into account the circular polarizations:
\begin{align}
\vec{\epsilon}(\vec{n}_{\mathcal{P}},\lambda=\pm1) &= \vec \epsilon_{\pm1} = \frac{1}{\sqrt 2}( \pm\vec e_1+i\vec e_2), \\
\vec{\epsilon}(\vec{n}_{\mathcal{P}},\lambda=0) &= \vec \epsilon_{0} = \vec{e}_{3},
\end{align}
and the helicity bispinors for this particular situation, $\xi_{-1/2}=(-1,0)^T$, $\xi_{1/2}=(1,0)^T$  and $\eta_{1/2}=(0,-1)^T$, we find that the only combinations of terms which surive the $\sigma, \lambda$ summation are:
\begin{align}
\xi^{\dag}_{-1/2}(\vec{p})\vec{\sigma}\cdot\vec{\epsilon}_{1}(\vec{n}_{\mathcal{P}},\lambda = \pm1)\eta_{1/2}(\vec{p}\,') &= \sqrt{2}, \\
\xi^{\dag}_{1/2}(\vec{p})\vec{\sigma}\cdot\vec{\epsilon}_{3}(\vec{n}_{\mathcal{P}},\lambda = 0)\eta_{1/2}(\vec{p}\,') &= 1.
\end{align}

The decay rate will be defined as the probability derivative with respect to time \cite{20}:
\begin{align}
P_{i \rightarrow f} &= \frac{1}{3} \sum_{\sigma} \left| \mathcal{A}_{[W^-\rightarrow e^- + \tilde{\nu}]}^{[\omega \rightarrow 0]} \right|^2,\nonumber\\
R_{i \rightarrow f} &= \frac{d P_{i \rightarrow f}}{dt} = \frac{1}{3} \sum_{\sigma} \frac{d \left| \mathcal{A}_{[W^-\rightarrow e^- + \tilde{\nu}]}^{[\omega \rightarrow 0]} \right|^2}{dt},
\end{align}
where the factor of $1/3$ comes from averaging the boson polarisations $\lambda = \pm 1, 0$.

The $\sigma$ sum of the squared amplitude gives an overall factor of $2$ to $R_{i \rightarrow f}$, that will come from the terms with bispinors.

The total decay rate will be obtained after an integration over the final momenta of the electron and antineutrino \cite{12,19,21}:
\begin{equation}
R_{[W^-\rightarrow e^- + \tilde{\nu}]}^{[\omega \rightarrow 0]} =\int d^3p\int d^3p'R_{i \rightarrow f}.
\end{equation}

By using formulas:
\begin{align}
|(2\pi)\delta(E+E'-\mathcal{E})|^2 &= (2\pi)\,t\,\delta(E+E'-\mathcal{E}), \\ 
|(2\pi)^3\delta^3(\vec{p}\,'+\vec{p}-\vec{P})|^2 &= (2\pi)^3V\delta^3(\vec{p}\,'+\vec{p}-\vec{P}),
\end{align}
the total decay rate per unit volume becomes:
\begin{align}\label{rataminkwdelta}
R_{[W^-\rightarrow e^- + \tilde{\nu}]}^{[\omega \rightarrow 0]} &= \frac{1}{3}\cdot\frac{g^2}{64(2\pi)^2}\int d^3p\int d^3p'\frac{1}{|\vec P|}\left(\frac{|\vec p|+|\vec p\,'|}{|\vec P|}+1\right)^2\nonumber\\
&\times\delta(E+E'-\mathcal{E})\delta^3(\vec{p}\,'+\vec{p}-\vec{P}).
\end{align} 

The momenta integrals can be solved by using methods from Minkowski field theory which allow us to transform the three dimensional integral with respect to $p$ into a four dimensional integral \cite{12,19,21,HGF}:
\begin{equation}
\int \frac{d^3 p}{2 E} = \int d^4 p \, \theta(E) \, \delta(p^2).
\end{equation}

In consequence the momenta integrals from \eqref{rataminkwdelta} can be solved further:
\begin{align}
I(p,p',E,E')&=\frac{1}{|\vec P|}\int d^3p\int d^3p'\,\frac{E}{2E}\delta^4(p+p'-P)\left(\frac{|\vec p\,|+|\vec p\,'|}{|\vec P|}+1\right)^2\nonumber\\
&=\frac{1}{|\vec P|}\int d^3p'\int d^4p\,E\,\delta^4(p+p'-P)\delta(p^2)\theta(E)\left(\frac{|\vec p\,|+|\vec p\,'|}{|\vec P|}+1\right)^2\nonumber\\
&=\frac{1}{|\vec P|}\int d^3p' (\mathcal{E}-E')\delta((P-p')^2)\theta(\mathcal{E}-E')\left(\frac{|\vec p\,'|+|\vec P-\vec p\,'|}{|\vec P|}+1\right)^2.
\end{align}

Momentum conservation and the fact that the momenta of the $W$ boson and antineutrino are all aligned in the direction of the third axis $\vec p=-p \vec e_3\,,\vec p\,'=p' \vec e_3\,,\vec P=P \vec e_3$, allow us to write $|\vec P-\vec p\,'|=|\vec P|-|\vec p\,'|$. This leads to:
\begin{align}
I(p,p',E,E') = \frac{4}{|\vec{P}|}  \int d\Omega_p'\int_0^\mathcal{E} dE'\,E'^2(\mathcal{E}-E') \delta(P^2 + p'^{\,2} - 2Pp'),
\end{align}
where the four-momentum relations for the antineutrino $p'^{\,2} = p'_{\mu}p'^{\,\mu} =0$ and for the boson momentum $P^2 = \mathcal{E}^2 - |\vec{P}|^2 = M_W^2$ hold.

In this case the delta factor reduces to:
\begin{align}
\delta((P-p')^2) = \frac{1}{2(\mathcal{E} -|\vec{P}|)} \delta\left(E' - \frac{\mathcal{E} +|\vec{P}|}{2}\right) .
\end{align}

The result for the $p'$ integral is obtained after performing the energy integral and the angular integral:
\begin{align}
I(p, p', E, E') &=\frac{4}{|\vec{P}|}\int d\Omega_p'\int_0^\mathcal{E} dE'\,E'^2(\mathcal{E}-E') \frac{1}{2(\mathcal{E} -|\vec{P}|)} \delta\left(E' - \frac{\mathcal{E} +|\vec{P}|}{2}\right)  \nonumber\\
&= \frac{\pi}{|\vec{P}|} \left(\mathcal{E}^2 + |\vec{P}|^2 + 2 \mathcal{E} |\vec{P}| \right),
\end{align}
where we have used equality $dp' p'^2 = dE' E'^2$. In the ultra-relativistic case when $\mathcal{E}\sim |\vec{P}|$, the final result is $I(p, p', E, E')=4 \pi \mathcal{E}$.

Let us recall the decay from Minkowski spacetime and consider the $W$ boson at rest. If we take $\mathcal{E}=2M_W$, the result is:
\begin{equation}
I(p, p', E, E') = 4\pi \mathcal{E} = 8 \pi M_W.
\end{equation}

The final formula for the decay rate is obtained by substituting the above result for the momenta integrals into formula \eqref{rataminkwdelta} and expressing the coupling constant in terms of the Fermi constant and the mass of the W bosons, $g^2=4\sqrt{2}G_FM_W^2$ \cite{rat}. 

The final result for the transition rate in the Minkowski limit is:
\begin{eqnarray}\label{rminkwfinal}
R_{[W^-\rightarrow e^- + \tilde{\nu}]}^{[\omega \rightarrow 0]}=\frac{G_FM_W^3}{6\pi\sqrt{2}}.
\end{eqnarray}

This result is in fact the well known formula from Minkowski spacetime \cite{rat}: 
\begin{equation}\label{decaminkw}
R_{\text{Minkowski}}=\frac{G_FM_W^3}{6\pi\sqrt{2}}=226\pm1 MeV, 
\end{equation}
for the decay $W^-\rightarrow e^- + \tilde{\nu}_{e}$. 

We have obtained the exact Minkowski limit of the transition rate for this process, starting from the amplitude in curved space-time. 

\section{\bf{$W^{\pm}$ boson emission by leptons}}

The second type of processes we are going to investigate is $W^{\pm}$ bosons being emitted by leptons. These types of processes are forbidden in Minkowski spacetime due to energy-momentum conservation. However, they are permitted in this de Sitter setup \cite{44}.

There are again six ways in which electrons and positrons can emit massive charged bosons:
\begin{align}
& e^{-} \rightarrow W^{-} + \nu_{e}  \quad\quad\quad\quad\,\quad  e^{+} \rightarrow W^{+} + \tilde{\nu}_e \nonumber\\
&\mu^{-} \rightarrow W^{-} + \nu_{\mu}  \quad\quad\quad\quad\quad  \mu^{+} \rightarrow W^{+} + \tilde{\nu}_\mu \nonumber\\
&\tau^{-} \rightarrow W^{-} + \nu_{\tau}  \quad\quad\quad\quad\,\quad \tau^{+} \rightarrow W^{+} + \tilde{\nu}_\tau \nonumber.
\end{align}

We use as an example the emission of a $W^{-}$ boson by an electron:
\begin{equation}
e^{-} \rightarrow W^{-} + \nu_{e}.
\end{equation}

The transition amplitude is then:
\begin{align}
\mathcal{A}_{[{e^{-} \rightarrow W^{-} + \nu}]} &= \frac{i g}{2\sqrt{2}} \int d^4x \sqrt{-g(x)}\,
\bar{u}_{\vec{p'}\sigma'}(x) \gamma^{\hat{\alpha}} e^{\mu}_{\hat{\alpha}}(1-\gamma^5)\, u_{\vec{p}\sigma}(x) f^{*}_{\vec{P}\lambda,_{\mu}}(x).
\end{align}

The solution for a neutrino of momentum  $\vec{p'}$ and helicity $\sigma'$ in de Sitter spacetime is \cite{22}:
\begin{align}\label{e:solnu}	
u_{\vec{p'},\sigma'}^{\dag}(t_{c},\vec{x})
=&  \left(\frac{-\omega t_{c}}{2\pi}\right)^{3/2} \Bigg(
\begin{array}{c}
\left(\frac{1}{2}-\sigma' \right)\xi_{\sigma'}^{\dag}(\vec{p'}), \,\,\,
0 
\end{array}\Bigg) e^{i p' t_{c}-i\vec{p'}\vec{x}},
\end{align}
where we have kept in mind that in the amplitude we have $\bar{u} = u^{\dag}\gamma^0$. 

For the electron we have a similar solution as before \cite{22}:
\begin{align}\label{e:sole}
u_{\vec{p},\sigma}(t,\vec{x})&= \frac{i}{(2\pi)^{3/2}}\sqrt{\frac{\pi p}{\omega}} \left(
\begin{array}{c}
\frac{1}{2}e^{k\frac{\pi}{2}}H_{\nu_{-}}^{(1)}(\frac{p}{\omega}e^{-\omega t})\xi_{\sigma}(\vec{p}) \\
\sigma e^{-k\frac{\pi}{2}}H_{\nu_{+}}^{(1)}(\frac{p}{\omega}e^{-\omega t})\xi_{\sigma}(\vec{p}) \\
\end{array}
\right)e^{i\vec{p}\vec{x}-2\omega t},
\end{align}
given with respect to the proper time $t$ and helicity spinors $\xi_{\sigma}(\vec{p})$.

For the Proca field we have the conjugate solutions, first for $\lambda=\pm 1$ \cite{2}:
\begin{align}\label{e:solProcaLambda1*}
f_{\vec{P}\lambda=\pm 1,_i}^{*}(x) &= 
\frac{\sqrt{\pi}e^{-\pi K/2}\sqrt{-t_c}}{2(2\pi)^{3/2}}H_{-iK}^{(2)}(-Pt_c) \epsilon_i(\vec{P},\lambda=\pm 1)e^{-i\vec{P}\vec{x}},
\end{align}
then for $\lambda=0$ we have the temporal component:
\begin{align}\label{e:solProcaLambda0t*}
f_{\vec{P}\lambda=0,_0}^{*}(x) &=
\frac{\sqrt{\pi}e^{-\pi K/2}\omega P (-t_c)^{3/2}}{2(2\pi)^{3/2}M_W} H_{-iK}^{(2)}(-Pt_c) e^{-i\vec{P}\vec{x}},
\end{align}
and the spatial one \cite{2}:
\begin{align}\label{e:solProcaLambda0s*}
f_{\vec{P}\lambda=0,_i}^{*}(x) &=
\frac{-i\sqrt{\pi}e^{-\pi K/2}\omega P}{2(2\pi)^{3/2}M_W}\,\epsilon_i(\vec{P},\lambda= 0) e^{-i\vec{P}\vec{x}}\nonumber \\
&\times\bigg[\left(\frac{1}{2}-iK\right)\frac{\sqrt{-t_c}}{P} H_{-iK}^{(2)}(-Pt_c)- (-t_c)^{3/2}H_{1-iK}^{(2)}(-Pt_c)\bigg],
\end{align}
given again as functions of the conformal time $t_c$ and polarisation vectors $\epsilon_i(\vec{P},\lambda)$.

\subsection{\bf{The amplitude for $\lambda=\pm 1$}}

The amplitude for this emission process is:
\begin{align}\label{AmpEmission}
\mathcal{A}_{[{ e^{-} \rightarrow W^{-}+ \nu}]}^{[\lambda = \pm 1]} &= \frac{i g}{2\sqrt{2}} \int d^4x \sqrt{-g(x)}\, \bar{u}_{\vec{p'}\sigma'}(x) \gamma^{\hat{\alpha}} e^{i}_{\hat{\alpha}}(1-\gamma^5)\, u_{\vec{p \sigma}}(x) f^{*}_{\vec{P}\lambda = \pm 1,_{i}}(x),
\end{align}
where we use only the Proca field solution for $\lambda = \pm 1$.

The steps for solving this amplitude are the same as in the previous section. We replace solutions \eqref{e:solnu},\eqref{e:sole} and \eqref{e:solProcaLambda1*} in amplitude \eqref{AmpEmission}. Thus we obtain the expression below:
\begin{align}\label{ampemissionlambda1hankel}
\mathcal{A}_{[{ e^{-} \rightarrow W^{-} + \nu}]}^{[\lambda=\pm 1]} &= \frac{- g}{4\sqrt{2}}\frac{\pi \sqrt{p} e^{\pi k/2} e^{-\pi K/2}}{(2\pi)^{3/2}}\left(\frac{1}{2}-\sigma' \right)\delta^3(\vec{P}+\vec{p'}-\vec{p}) \nonumber\\
&\times \int z\, dz\, e^{-ip' z}H_{\frac{1}{2}-ik}^{(1)}(pz)H_{-iK}^{(2)}(Pz)\xi_{\sigma'}^{\dag}(\vec{p'})\sigma^{j}\xi_{\sigma}(\vec{p})\epsilon_{j}(\vec{P},\lambda=\pm 1), 
\end{align}
where $z=-t_c=e^{-\omega t}/\omega$.

The exponential and the Hankel functions in the integral above will be transformed into Bessel J and K functions with the help of formulas \eqref{exptok} - \eqref{hankeltoj} from \hyperref[AppendixA]{Appendix A} \cite{AS,21}. This will help us split the integral with respect to $z$ into four others: 
\begin{align}
T_1 =& \int dz\, z\sqrt{z}\, i e^{-\pi k } J_{\frac{1}{2}}(p'z) J_{\frac{1}{2} -ik}(pz)  K_{-iK}(iPz), \nonumber\\
T_2 =& \int dz\, z\sqrt{z}\, J_{\frac{1}{2}}(p'z) J_{-\frac{1}{2} +ik}(pz)  K_{-iK}(iPz), \nonumber\\
T_3 =& \int dz\, z\sqrt{z}\, e^{-\pi k } J_{-\frac{1}{2}}(p'z) J_{\frac{1}{2} -ik}(pz)  K_{-iK}(iPz), \nonumber\\
T_4 =& \int dz\, z\sqrt{z}\, i^{-1} J_{-\frac{1}{2}}(p'z) J_{-\frac{1}{2} +ik}(pz)  K_{-iK}(iPz).
\end{align}

Consequently the amplitude will take the following form:
\begin{align}\label{AmpEmissionLambda1}
\mathcal{A}_{[{ e^{-} \rightarrow W^{-} + \nu}]}^{[\lambda=\pm 1]} &= -\frac{g\sqrt{i}}{4}\frac{\sqrt{\pi}\sqrt{p p'}e^{\pi k/2}e^{-i \pi/4}}{(2\pi)^{3/2}i\cosh{(\pi k)}} \left(\frac{1}{2}-\sigma'\right)\delta^3(\vec{P}+\vec{p'}-\vec{p})\nonumber\\
&\times \left[T_1 + T_2 - T_3 - T_4\right]\xi_{\sigma'}^{\dag}(\vec{p'})\sigma^{j}\xi_{\sigma}(\vec{p})\epsilon_{j}(\vec{P},\lambda=\pm 1),
\end{align}
where terms $T_1 - T_4$ can be integrated using formula \eqref{intBesseltoF4} from \hyperref[AppendixA]{Appendix A} \cite{AS,21}:
\begin{align}\label{t1}
T_1 &= \frac{i e^{-\pi k}\sqrt{2} p'^{\frac{1}{2}} p^{\frac{1}{2}-ik} (iP)^{-\frac{7}{2}+ik}}{\Gamma\left(\frac{3}{2}\right)\Gamma\left(\frac{3}{2}-ik\right)}\, \Gamma\left(\frac{\frac{7}{2}-i(k-K)}{2}\right)  \Gamma\left(\frac{\frac{7}{2}-i(k+K)}{2}\right)\nonumber \\
&\times F_{4}\Bigg( \frac{\frac{7}{2}-i(k-K)}{2}, \frac{\frac{7}{2}-i(k+K)}{2}, \frac{3}{2}, \frac{3}{2} -ik,\frac{p'^2}{P^2},\frac{p^2}{P^2}\Bigg),
\end{align}

\begin{align}\label{t2}
T_2 &= \frac{\sqrt{2} p'^{\frac{1}{2}} p^{-\frac{1}{2}+ik} (iP)^{-\frac{5}{2}-ik}}{\Gamma\left(\frac{3}{2}\right)\Gamma\left(\frac{1}{2}+ik\right)} \, \Gamma\left(\frac{\frac{5}{2}+i(k+K)}{2}\right)  \Gamma\left(\frac{\frac{5}{2}+i(k-K)}{2}\right)\nonumber \\
&\times F_{4}\Bigg(\frac{\frac{5}{2}+i(k+K)}{2}, \frac{\frac{5}{2}+i(k-K)}{2},\frac{3}{2}, \frac{1}{2} + ik,\frac{p'^2}{P^2},\frac{p^2}{P^2}\Bigg),
\end{align}

\begin{align}\label{t3}
T_3 &= \frac{e^{-\pi k}\sqrt{2} p'^{-\frac{1}{2}} p^{\frac{1}{2}-ik} (iP)^{-\frac{5}{2}+ik}}{\Gamma\left(\frac{1}{2}\right)\Gamma\left(\frac{3}{2}-ik\right)} \, \Gamma\left(\frac{\frac{5}{2}-i(k-K)}{2}\right)  \Gamma\left(\frac{\frac{5}{2}-i(k+K)}{2}\right)\nonumber \\
&\times F_{4}\Bigg(\frac{\frac{5}{2}-i(k-K)}{2}, \frac{\frac{5}{2}-i(k+K)}{2},\frac{1}{2}, \frac{3}{2} - ik,\frac{p'^2}{P^2},\frac{p^2}{P^2}\Bigg),
\end{align}

\begin{align}\label{t4}
T_4 &= \frac{i^{-1} \sqrt{2} p'^{-\frac{1}{2}} p^{-\frac{1}{2}+ik} (iP)^{-\frac{3}{2}-ik}}{\Gamma\left(\frac{1}{2}\right)\Gamma\left(\frac{1}{2}+ik\right)} \, \Gamma\left(\frac{\frac{3}{2}+i(k+K)}{2}\right)  \Gamma\left(\frac{\frac{3}{2}+i(k-K)}{2}\right)\nonumber \\
&\times F_{4}\Bigg(\frac{\frac{3}{2}+i(k+K)}{2}, \frac{\frac{3}{2}+i(k-K)}{2}, \frac{1}{2},\frac{1}{2} +ik,\frac{p'^2}{P^2},\frac{p^2}{P^2}\Bigg),
\end{align}
where Appell's function $F_4$ is defined in formula \eqref{F4series} from \hyperref[AppendixA]{Appendix A} \cite{AS,21}.

\subsection{\bf{The amplitude for $\lambda = 0 $}}

We move forward to compute the amplitude for polarisation $\lambda = 0$. Since in this case we have both temporal \eqref{e:solProcaLambda0t*} and spatial \eqref{e:solProcaLambda0s*} components of the Proca solution, we split the amplitude into two integrals:
\begin{align}
\mathcal{A}_{[{e^{-} \rightarrow W^{-} +  \nu}]}^{[\lambda = 0]} &= \frac{i g}{2\sqrt{2}} \int d^4x \sqrt{-g(x)} \bar{u}_{\vec{p'}\sigma'}(x) \gamma^{\hat{\alpha}} e^{0}_{\hat{\alpha}}(1-\gamma^5) u_{\vec{p} \sigma}(x) f^{*}_{\vec{P}\lambda = 0,_{0}}(x) \nonumber\\
&+ \frac{i g}{2\sqrt{2}} \int d^4x \sqrt{-g(x)}\bar{u}_{\vec{p'}\sigma'}(x) \gamma^{\hat{\alpha}} e^{i}_{\hat{\alpha}}(1-\gamma^5) u_{\vec{p} \sigma}(x) f^{*}_{\vec{P}\lambda = 0,_{i}}(x). \nonumber\\
\end{align}

Then we further split the amplitude into three terms:
\begin{align}\label{ampelambda0123}
\mathcal{A}_{[{e^{-} \rightarrow W^{-} +  \nu}]}^{[\lambda = 0]} &= \,\,\mathcal{A}_{[{e^{-} \rightarrow W^{-} +  \nu}]}^{[\lambda = 0](1)} + \mathcal{A}_{[{e^{-} \rightarrow W^{-} +  \nu}]}^{[\lambda = 0](2)} +\mathcal{A}_{[{e^{-} \rightarrow W^{-} +  \nu}]}^{[\lambda = 0](3)},
\end{align}
where the first term contains the temporal solution \eqref{e:solProcaLambda0t*}, while terms two and three account for the two parts of the spatial Proca solution \eqref{e:solProcaLambda0s*} \cite{2}.

We solve first the amplitude for the temporal component of the Proca equation:
\begin{align}
\mathcal{A}^{[\lambda=0](1)}_{[{e^{-} \rightarrow W^{-} +  \nu}]} &= \frac{g}{4\sqrt{2}}\frac{\pi P\sqrt{p} e^{\pi k /2}e^{-\pi K/2}\omega}{\left(2\pi\right)^{3/2}M_W} \left(\frac{1}{2} - \sigma'\right) \delta^3\left(\vec{P} + \vec{p'} - \vec{p}\right)  \nonumber \\
&\times\int z^2 \, dz\, e^{-i p' z} H_{1/2-ik}^{(1)}(pz)H_{-iK}^{(2)}(Pz)\,\xi^{\dag}_{\sigma'}(\vec{p'}) \xi_{\sigma}(\vec{p}).
\end{align}

In terms of Bessel functions we arrive at:
\begin{align}\label{ampelambda01}
\mathcal{A}^{[\lambda=0](1)}_{[{e^{-} \rightarrow W^{-} +  \nu}]} &= \frac{g\sqrt{i}}{4}\frac{\omega}{M_W}\frac{P\sqrt{\pi}\sqrt{p p'}e^{\pi k/2}e^{-i \pi/4}}{(2\pi)^{3/2}\cosh{(\pi k)}} \left(\frac{1}{2}-\sigma'\right)\delta^3(\vec{P}+\vec{p'}-\vec{p})\nonumber\\
&\times \left[T_5 - T_6 - T_7 + T_8\right]\,\xi_{\sigma'}^{\dag}(\vec{p'})\xi_{\sigma}(\vec{p}),
\end{align}
where we have the following integrals:
\begin{align}
T_5 =& \int dz\, z^2\sqrt{z}\, i e^{-\pi k } J_{\frac{1}{2}}(p'z) J_{\frac{1}{2} -ik}(pz)  K_{-iK}(iPz), \nonumber\\
T_6 =& \int dz\, z^2\sqrt{z}\, J_{\frac{1}{2}}(p'z) J_{-\frac{1}{2} +ik}(pz)  K_{-iK}(iPz), \nonumber\\
T_7 =& \int dz\, z^2\sqrt{z}\, e^{-\pi k } J_{-\frac{1}{2}}(p'z) J_{\frac{1}{2} -ik}(pz)  K_{-iK}(iPz), \nonumber\\
T_8 =& \int dz\, z^2\sqrt{z}\, i^{-1} J_{-\frac{1}{2}}(p'z) J_{-\frac{1}{2} +ik}(pz)  K_{-iK}(iPz).
\end{align}

By applying formula \eqref{intBesseltoF4} to the above terms we obtain \cite{AS,21}:
\begin{align}
T_5 &= \frac{i e^{-\pi k} 2^{3/2} p'^{\frac{1}{2}} p^{\frac{1}{2}-ik} (iP)^{-\frac{9}{2}+ik}}{\Gamma\left(\frac{3}{2}\right)\Gamma\left(\frac{3}{2}-ik\right)} \, \Gamma\left(\frac{\frac{9}{2}-i(k-K)}{2}\right)  \Gamma\left(\frac{\frac{9}{2}-i(k+K)}{2}\right)\nonumber \\
&\times F_{4}\Bigg( \frac{\frac{9}{2}-i(k-K)}{2}, \frac{\frac{9}{2}-i(k+K)}{2}, \frac{3}{2}, \frac{3}{2} - ik,\frac{p'^2}{P^2},\frac{p^2}{P^2}\Bigg),
\end{align}

\begin{align}
T_6 &= \frac{2^{3/2} p'^{\frac{1}{2}} p^{-\frac{1}{2}+ik} (iP)^{-\frac{7}{2}-ik}}{\Gamma\left(\frac{3}{2}\right)\Gamma\left(\frac{1}{2}+ik\right)} \, \Gamma\left(\frac{\frac{7}{2}+i(k+K)}{2}\right)  \Gamma\left(\frac{\frac{7}{2}+i(k-K)}{2}\right)\nonumber \\
&\times F_{4}\Bigg(\frac{\frac{7}{2}+i(k+K)}{2}, \frac{\frac{7}{2}+i(k-K)}{2},\frac{3}{2}, \frac{1}{2} + i k,\frac{p'^2}{P^2},\frac{p^2}{P^2}\Bigg),
\end{align}

\begin{align}
T_7 &= \frac{e^{-\pi k}2^{3/2} p'^{-\frac{1}{2}} p^{\frac{1}{2}-ik} (iP)^{-\frac{7}{2}+ik}}{\Gamma\left(\frac{1}{2}\right)\Gamma\left(\frac{3}{2}-ik\right)} \, \Gamma\left(\frac{\frac{7}{2}-i(k-K)}{2}\right)  \Gamma\left(\frac{\frac{7}{2}-i(k+K)}{2}\right)\nonumber \\
&\times F_{4}\Bigg(\frac{\frac{7}{2}-i(k-K)}{2}, \frac{\frac{7}{2}-i(k+K)}{2}, \frac{1}{2}, \frac{3}{2} - ik,\frac{p'^2}{P^2},\frac{p^2}{P^2}\Bigg),
\end{align}

\begin{align}
T_8 &= \frac{i^{-1} 2^{3/2} p'^{-\frac{1}{2}} p^{-\frac{1}{2}+ik} (iP)^{-\frac{5}{2}-ik}}{\Gamma\left(\frac{1}{2}\right)\Gamma\left(\frac{1}{2}+ik\right)} \, \Gamma\left(\frac{\frac{5}{2}+i(k+K)}{2}\right)  \Gamma\left(\frac{\frac{5}{2}+i(k-K)}{2}\right)\nonumber \\
&\times F_{4}\Bigg(\frac{\frac{5}{2}+i(k+K)}{2}, \frac{\frac{5}{2}+i(k-K)}{2},\frac{1}{2}, \frac{1}{2} + ik,\frac{p'^2}{P^2},\frac{p^2}{P^2}\Bigg).
\end{align}

Computing for $A^{(2)}$ one can see that this term is actually very similar to the amplitude for $\lambda=\pm 1$ \eqref{AmpEmissionLambda1}, up to an imaginary factor and the ratio $\omega/M_W$, but with a different polarisation vector: 
\begin{align}\label{ampWlambda02el}
\mathcal{A}^{[\lambda=0](2)}_{[{e^{-} \rightarrow W^{-} +  \nu}]} = - \frac{i \,\omega}{M_W} \left(\frac{1}{2} - iK\right)  \mathcal{A}(\pm1) \xi_{\sigma'}^{\dag}(\vec{p'})\sigma^{j}\xi_{\sigma}(\vec{p})\epsilon_{j}(\vec{P},\lambda=0),
\end{align}
where we denote:
\begin{align}
\mathcal{A}(\pm 1) &= -\frac{g\sqrt{i}}{4}\frac{\sqrt{\pi}\sqrt{p p'}e^{\pi k/2}e^{-i \pi/4}}{(2\pi)^{3/2}i\cosh{(\pi k)}} \left(\frac{1}{2}-\sigma'\right)\delta^3(\vec{P}+\vec{p'}-\vec{p})\nonumber\\
&\times \left[T_1 + T_2 - T_3 - T_4\right],
\end{align}
since we can rewrite the amplitude for $\lambda=\pm1$ \eqref{AmpEmissionLambda1} as such:
\begin{align}
\mathcal{A}_{[{e^{-} \rightarrow W^{-} +  \nu}]}^{[\lambda=\pm 1]} &= \mathcal{A}(\pm1) \xi_{\sigma'}^{\dag}(\vec{p'})\sigma^{j}\xi_{\sigma}(\vec{p})\epsilon_{j}(\vec{P},\lambda=\pm 1).
\end{align}

We are left with the third term:
\begin{align}\
\mathcal{A}^{[\lambda=0](3)}_{[{e^{-} \rightarrow W^{-} +  \nu}]} &= -\frac{i g}{4\sqrt{2}}\frac{\pi P\sqrt{p}  e^{\pi k /2}e^{-\pi K/2}\omega}{\left(2\pi\right)^{3/2}M_W} \left(\frac{1}{2} - \sigma'\right) \delta^3\left(\vec{P} + \vec{p'} - \vec{p}\right)  \nonumber \\
&\times\int z^2 \, dz\, e^{-i p' z} H_{1/2-ik}^{(1)}(pz)H_{1-iK}^{(2)}(Pz)\nonumber\\
&\times\xi_{\sigma'}^{\dag}(\vec{p'})\sigma^{j}\xi_{\sigma}(\vec{p})\epsilon_{j}(\vec{P},\lambda=0),
\end{align}
which can also be written as:
\begin{align}\label{ampelambda03}
\mathcal{A}^{[\lambda=0](3)}_{[{W^{-} \rightarrow e^{-} + \nu}]} &= -\frac{i \sqrt{i}g}{4}\frac{P\sqrt{\pi}\sqrt{p p'}  e^{-i\pi/4} e^{\pi k /2} \omega}{\left(2\pi\right)^{3/2}M_W \cosh{(\pi k)}} \left(\frac{1}{2} - \sigma'\right) \delta^3\left(\vec{P} + \vec{p'} - \vec{p}\right)  \nonumber \\
&\times\left[T_9 + T_{10} - T_{11} - T_{12} \right] \,\xi_{\sigma'}^{\dag}(\vec{p'})\sigma^{j}\xi_{\sigma}(\vec{p})\epsilon_{j}(\vec{P},\lambda= 0).
\end{align}

Terms $T_9 - T_{12}$ are defined below:
\begin{align}
T_9 =& \int dz\, z^2\sqrt{z}\, i e^{-\pi k } J_{\frac{1}{2}}(p'z) J_{\frac{1}{2} -ik}(pz)  K_{1-iK}(iPz), \nonumber\\
T_{10} =& \int dz\, z^2\sqrt{z}\, J_{\frac{1}{2}}(p'z) J_{-\frac{1}{2} +ik}(pz)  K_{1-iK}(iPz), \nonumber\\
T_{11} =& \int dz\, z^2\sqrt{z}\, e^{-\pi k } J_{-\frac{1}{2}}(p'z) J_{\frac{1}{2} -ik}(pz)  K_{1-iK}(iPz), \nonumber\\
T_{12} =& \int dz\, z^2\sqrt{z}\, i^{-1} J_{-\frac{1}{2}}(p'z) J_{-\frac{1}{2} +ik}(pz)  K_{1-iK}(iPz).
\end{align}

After solving the integrals the terms become:
\begin{align}
T_9 &= \frac{i e^{-\pi k} 2^{3/2} p'^{\frac{1}{2}} p^{\frac{1}{2} - ik} (iP)^{-\frac{9}{2} + ik}}{\Gamma\left(\frac{3}{2}\right)\Gamma\left(\frac{3}{2}-ik\right)} \, \Gamma\left(\frac{\frac{7}{2}-i(k-K)}{2}\right)  \Gamma\left(\frac{\frac{11}{2}-i(k+K)}{2}\right)\nonumber \\
&\times F_{4}\Bigg( \frac{\frac{7}{2}-i(k-K)}{2}, \frac{\frac{11}{2}-i(k+K)}{2},\frac{3}{2}, \frac{3}{2} - ik,\frac{p'^2}{P^2},\frac{p^2}{P^2}\Bigg),
\end{align}

\begin{align}
T_{10} &= \frac{2^{3/2} p'^{\frac{1}{2}} p^{-\frac{1}{2}+ik} (iP)^{-\frac{7}{2}-ik}}{\Gamma\left(\frac{3}{2}\right)\Gamma\left(\frac{1}{2}+ik\right)} \, \Gamma\left(\frac{\frac{5}{2}+i(k+K)}{2}\right)  \Gamma\left(\frac{\frac{9}{2}+i(k-K)}{2}\right)\nonumber \\
&\times F_{4}\Bigg(\frac{\frac{5}{2}+i(k+K)}{2}, \frac{\frac{9}{2}+i(k-K)}{2},\frac{3}{2}, \frac{1}{2} + i k,\frac{p'^2}{P^2},\frac{p^2}{P^2}\Bigg),
\end{align}

\begin{align}
T_{11} &= \frac{e^{-\pi k}2^{3/2} p'^{-\frac{1}{2}} p^{\frac{1}{2}-ik} (iP)^{-\frac{7}{2}+ik}}{\Gamma\left(\frac{1}{2}\right)\Gamma\left(\frac{3}{2}-ik\right)} \, \Gamma\left(\frac{\frac{5}{2}-i(k-K)}{2}\right)  \Gamma\left(\frac{\frac{9}{2}-i(k+K)}{2}\right)\nonumber \\
&\times F_{4}\Bigg(\frac{\frac{5}{2}-i(k-K)}{2}, \frac{\frac{9}{2}-i(k+K)}{2}, \frac{1}{2}, \frac{3}{2} - ik,\frac{p'^2}{P^2},\frac{p^2}{P^2}\Bigg),
\end{align}

\begin{align}
T_{12} &= \frac{i^{-1} 2^{3/2} p'^{-\frac{1}{2}} p^{-\frac{1}{2}+ik} (iP)^{-\frac{5}{2}-ik}}{\Gamma\left(\frac{1}{2}\right)\Gamma\left(\frac{1}{2}+ik\right)} \, \Gamma\left(\frac{\frac{3}{2}+i(k+K)}{2}\right)  \Gamma\left(\frac{\frac{7}{2}+i(k-K)}{2}\right)\nonumber \\
&\times F_{4}\Bigg(\frac{\frac{3}{2}+i(k+K)}{2}, \frac{\frac{7}{2}+i(k-K)}{2},\frac{1}{2}, \frac{1}{2} + ik,\frac{p'^2}{P^2},\frac{p^2}{P^2}\Bigg).
\end{align}

As was the case in the previous process as well, all terms \eqref{ampelambda01}, \eqref{ampWlambda02el} and \eqref{ampelambda03} that comprise amplitude \eqref{ampelambda0123} for $\lambda=0$ have a factor of $\omega/M_W$. Only the second term \eqref{ampWlambda02el} survies when $\omega \rightarrow 0$, the same as in the previous process. Therefore we can assert that, for longitudinal polarisation, only the term which is proportional with the transversal polarisation amplitude survives in the Minkowski limit.

\subsection{\bf{The transition rate}}

First we will compute the rate in as general a case as we possibly can, i.e. we will keep some dependence on the ratios between the particle masses and Hubble parameter through variables $k$ and $K$. 

We rewrite amplitude \eqref{ampemissionlambda1hankel} for $\lambda=\pm 1$ in order to change the Hankel function corresponding to momentum $P$ of the $W^{-}$ boson into a Bessel K function, by using the relation \eqref{hankeltok}:
\begin{align}\label{ampemissionratahk}
\mathcal{A}_{[{ e^{-} \rightarrow W^{-} + \nu}]}^{[\lambda=\pm 1]} &= \frac{-g e^{\pi k/2}}{4\sqrt{2}}\frac{\pi \sqrt{p}  }{(2\pi)^{3/2}}\,\delta^3(\vec{P}+\vec{p'}-\vec{p})\frac{2i}{\pi} \nonumber\\
&\times \int z\, dz\, e^{-ip' z}H^{(1)}_{\frac{1}{2}-ik}(pz)K_{-iK}(iPz)\,\xi_{-1/2}^{\dag}(\vec{p'})\sigma^{j}\xi_{\sigma}(\vec{p})\epsilon_{j}(\vec{P},\lambda=\pm 1),
\end{align}
where we have taken into account the fact that the helicity of the neutrino has to be $\sigma' = -1/2$.

The amplitude has the following general form:
\begin{equation}\label{dsrt}
\mathcal{A}_{[{ e^{-} \rightarrow W^{-} + \nu}]}^{[\lambda=\pm 1]} = \delta^3(\vec p_f-\vec p _i)\mathcal{M}_{if} \mathcal{I}_{if},
\end{equation}
where the delta function assures momentum conservation in this the process. 

We denote the temporal integral by $\mathcal{I}_{if}$:
\begin{equation}
\mathcal{I}_{if}=\int_0^{\infty} dt \, \mathcal{K}_{if},
\end{equation}
where $\mathcal{K}_{if}$ is the integrand of the temporal integral:
\begin{equation}\label{kifrataemisie}
\mathcal{K}_{if} = z\, dz\, e^{-ip' z}H^{(1)}_{\frac{1}{2}-ik}(pz)K_{-iK}(iPz),
\end{equation}
while $\mathcal{M}_{if} $ denotes the rest of the factors in the amplitude:
\begin{align}
\mathcal{M}_{if} &= -\frac{i g}{2\sqrt{2}} \frac{\sqrt{p} e^{\pi k /2}}{(2\pi)^{3/2}} \left(\frac{1}{2} -\sigma'\right)\, \xi_{-1/2}^{\dag}(\vec{p'})\sigma^{j}\xi_{\sigma}(\vec{p})\epsilon_{j}(\vec{P},\lambda=\pm 1).
\end{align}

The transition rate is defined as follows \cite{cpc}:
\begin{align}\label{rt}
R_{i\rightarrow f} &= \frac{1}{2} \sum_{\sigma'\lambda}\frac{1}{(2\pi)^3}\,\delta^3(\vec{p}\,'+\vec{P}-\vec{p}\,)| \mathcal{M}_{if}|^2  | \mathcal{I}_{if}| \lim_{t\rightarrow \infty}| e^{\omega t}\mathcal{K}_{if}|.
\end{align}

The transition rate is relevant when $\omega\gg M_{W}$ and $\omega\gg m_{e}$, so we can further approximate the index of the modified Bessel function $K_{-iK}(z)$ from \eqref{ampemissionratahk} such that:
\begin{equation}
-i\sqrt{\left(\frac{M_{W}}{\omega}\right)^{2}-\frac{1}{4}} = \frac{1}{2}\sqrt{1 - 4\left(\frac{M_{W}}{\omega}\right)^{2}}\approx \frac{1}{2}. 
\end{equation}

An analytical result for the transition rate can be obtained if the Bessel K functions are expanded for small arguments \cite{AS,21}:
\begin{align}
K_{\nu} (z) \simeq \frac{2^{\nu -1} \Gamma(\nu)}{z^\nu}, 
\end{align}
since for $t\rightarrow \infty$, the argument becomes very small, that is $z=e^{-\omega t}/\omega \rightarrow 0$. 

Then the Bessel K function reduces to:
\begin{equation}
K_{1/2} \left(\frac{Pe^{-\omega t}}{\omega}\right) \simeq \frac{\Gamma(1/2)}{\sqrt{2i\frac{Pe^{-\omega t}}{\omega}}}.
\end{equation}

The Hankel function $H^{(1)}_{1/2-ik}(pz)$, can also be approximated for small arguments \cite{AS,21}:
\begin{equation*}
H^{(1)}_{\nu}(z)\simeq -i\frac{2^{\nu} \Gamma(\nu)}{\pi z^\nu},
\end{equation*}
which in our case translates to the following formula:
\begin{equation}
H^{(1)}_{1/2-ik} \left(\frac{pe^{-\omega t}}{\omega}\right)\simeq \frac{-i\Gamma(1/2-ik)}{\pi}\left(\frac{2\omega}{pe^{-\omega t}}\right)^{1/2-ik}.
\end{equation}

Thus the limit of $\mathcal{K}_{i \rightarrow f}$ \eqref{kifrataemisie} can be computed by using the above approximated functions, which still preserve the dependence on parameter $k=m_e/\omega$, keeping in mind that $z=e^{-\omega t}/\omega$:
\begin{equation}\label{lim}
\lim_{t \rightarrow t_{\infty}} \big| e^{\omega t} \mathcal{K}_{i \rightarrow f} \big| = \frac{1}{ \sqrt{p P \cosh(\pi k)}},
\end{equation}
where by $t_{\infty}$ we denote a sufficiently large time after the interaction.

With this result for the limit we can write down the final expression for the transition rate \eqref{rt}:
\begin{align}
R_{i\rightarrow f} &= \frac{1}{2}\sum_{\lambda} \frac{g^2\pi^2pp'}{32(2\pi)^{3}}\delta^{3}(\vec{p}-\vec{p^{'}} - \vec{P})\,\frac{|\mathcal{I}_{if}| }{\sqrt{Pp\cosh(\pi k)}}\nonumber\\
&\times |\xi_{-1/2}^{\dag}(\vec{p^{'}})\sigma^{i}\xi_{\sigma}(\vec{p}){\epsilon}^{*}_{i}(\vec{P},\lambda= \pm1)|^{2} ,
\end{align}
where $|\mathcal{I}_{if}| = \sqrt{\mathcal{I}_{if}\mathcal{I}_{if}^{*}}$.

The expression for $\mathcal{I}_{if}$ is:
\begin{equation}
\mathcal{I}_{if}=\frac{e^{i\pi/4}e^{\pi k/2}}{i\sqrt{p'}\cosh(\pi k)}(T_1 + T_2 - T_3 - T_4),
\end{equation}
where terms $T_1$ to $T_4$ are defined in equations \eqref{t1}, \eqref{t2}, \eqref{t3}, \eqref{t4}.

For the sake of examinig the behaviour of the transition rate it is useful to perform a graphical analysis of the terms which depend on the ratios of the particle masses and the Hubble parameter, which we group into:
\begin{align}\label{Rgraff}
\mathcal{R} &= \frac{\sqrt{p p'} e^{\pi k/2}}{\cosh{(\pi k)}}  \sqrt{\frac{(T_1 + T_2 - T_3 - T_4)(T_1 + T_2 - T_3 - T_4)^{*}}{P \cosh{(\pi k)}}}.
\end{align}

\begin{figure}[H]
	\centering
	\includegraphics[width=7.9cm]{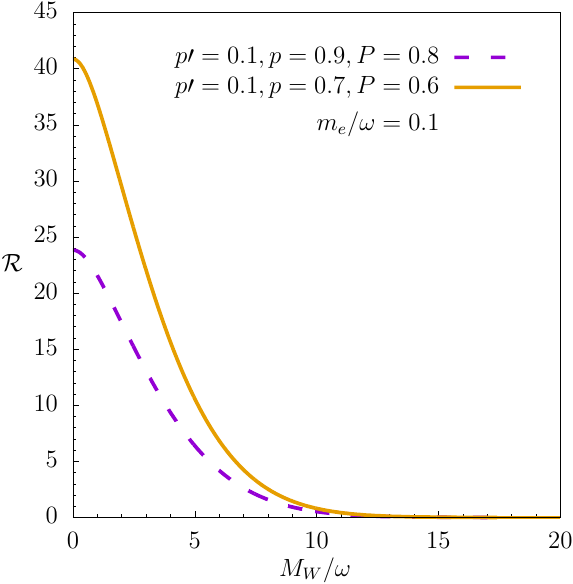}
	\caption{Transition rate $\mathcal{R}$ \eqref{Rgraff} as a function of $M_W/\omega$, for given values of the momenta and a fixed value of $m_e/\omega$.}
	\label{fig.1}
\end{figure}

\begin{figure}[H]
	\centering
	\includegraphics[width=7.9cm]{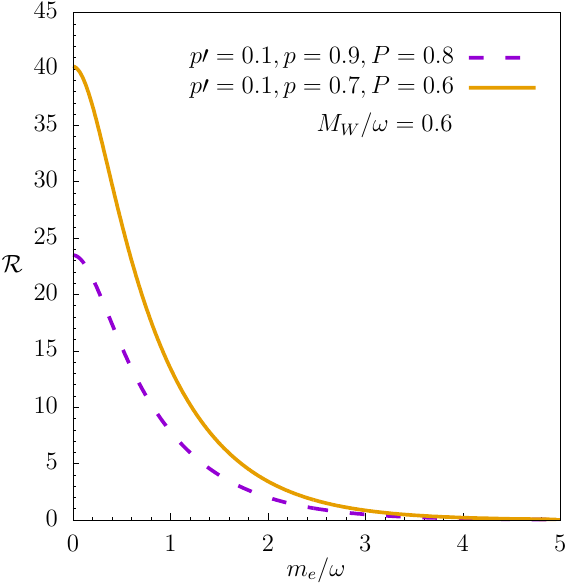}
	\caption{Transition rate $\mathcal{R}$ \eqref{Rgraff} as a function of $m_e/\omega$, for given values of the momenta and a fixed value of $M_W/\omega$.}
	\label{fig.2}
\end{figure}

The graphical analysis in Fig. \ref{fig.1} and Fig. \ref{fig.2} proves that the transition rate is significant as long as the ratio between the particles masses and the expansion parameter is small. The rate begins to decrease as the ratio increases. This result confirms that this process could be possible only in the early Universe, when $M_W/\omega \rightarrow 0$ and $m_e/\omega \rightarrow 0$. For larger values of the ration between particles masses and Hubble parameter, the transition rates vanish, confirming that when there is no spacetime expansion and $\omega \rightarrow 0$, the rates for $W$ bosons emmitted by leptons are zero. We indeed recover the Minkowski limit, when processes where $W$ bosons could be generated by fermions are forbidden due to energy-momentum conservation.

\subsection{\bf{The transition rate in the large expansion limit}}

In the limit when the expansion parameter is much larger than the particles masses $m_e/\omega=M_W/\omega=0$, the amplitude assumes a simpler form:
\begin{align}
\mathcal{A}_{[{ e^{-} \rightarrow W^{-} + \nu}]}^{[\lambda=\pm 1]} &= \frac{-ge^{i\pi/4}}{4\sqrt{2}}\frac{ \sqrt{p}}{(2\pi)^{3/2}}\,\xi_{\sigma'}^{\dag}(\vec{p'})\sigma^{j}\xi_{\sigma}(\vec{p})\epsilon_{j}(\vec{P},\lambda=\pm 1)\nonumber\\
&\times\left(\frac{1}{2}-\sigma' \right)\delta^3(\vec{P}+\vec{p'}-\vec{p})\, \frac{2}{\pi\sqrt{pP}}\int  dz\, e^{i(p-p'-P)  z},
\end{align}
since the Hankel functions reduce to exponential functions \eqref{exptok}.

We can again split the amplitude into factors to separate the temporal integral:
\begin{equation}
\mathcal{A}_{[{ e^{-} \rightarrow W^{-} + \nu}]}^{[\lambda=\pm 1]}  = \delta^3(\vec{P}+\vec{p'}-\vec{p})\mathcal{M}_{if}\mathcal{I}_{if},  
\end{equation}
where $\mathcal{I}_{if}$ is given by:
\begin{equation}
\mathcal{I}_{if}=\frac{2}{\pi\sqrt{pP}}\int  dz\, e^{i(p-p'-P)  z}=\frac{2i}{\pi\sqrt{pP}(p-p'-P)}.
\end{equation}

The limit of the integrand of temporal integral is:
\begin{align}
\lim_{t \rightarrow t_{\infty}} \big| e^{\omega t} \mathcal{K}_{i \rightarrow f} \big| =\frac{2}{\pi\sqrt{pP}}.
\end{align}

The transition rate will be defined using equation (\ref{rt}): 
\begin{align}
R_{i\rightarrow f}=\frac{g^2 \delta^3(\vec{P}+\vec{p'}-\vec{p})}{8(2\pi)^6P(p-p'-P)},
\end{align}
where we have considered the particles in this process to be aligned in the same direction, i.e. $\vec p = p \,\vec e_3\,,\vec p\,' =- p' \vec e_3\,,\vec P = P \vec e_3$.

The total transition rate for the $W$ boson emission by electrons is obtained after we integrate over the final momenta of the particles:
\begin{align}
R_{{ e^{-} \rightarrow W^{-} + \nu}} &= \int d^3 p'\int d^3P\,R_{i\rightarrow f}\nonumber\\
&= \frac{g^2}{8(2\pi)^6}\int d^3 p'\int d^3P\frac{\delta^3(\vec{P}+\vec{p'}-\vec{p})}{P(p-p'-P)}.
\end{align}

The first integration over momentum $p'$ leads to:
\begin{align}
I  &=\frac{1}{(2\pi)^3}\int d^3 p'\int d^3P\frac{\delta^3(\vec{P}+\vec{p'}-\vec{p})}{P(p-p'-P)}\nonumber\\
&=-\frac{1}{(2\pi)^3}\int d^3 p'\frac{1}{2p'(p+p')}
\end{align}

We can renormalize the integral by using dimensional regularization and the minimal subtraction method \cite{WF,GHV,IT,GT,HGF}. In $D$ dimensions this integral reads:
\begin{align}
I(D) &=-\frac{1}{(2\pi)^D}\int d^D p'\frac{1}{2p'(p+p')}\nonumber\\
&=-\frac{2\pi^{D/2}}{2(2\pi)^D\Gamma(D/2)}\int d p'\frac{p'^{D-2}}{(p+p')}.
\end{align}

The integral with respcet to $p'$ can be solved by using the Beta Euler function \eqref{betaeuler}, after performing a change of variables such that $p'=py$:
\begin{align}
I(D)  &=-\frac{2\pi^{D/2}p^{D-2}}{2(2\pi)^D\Gamma(D/2)}\int dy \frac{y^{D-2}}{(1+y)}\nonumber\\
&=-\frac{2\pi^{D/2}p^{D-2}\Gamma(D-1)\Gamma(4-D)}{2(2\pi)^D\Gamma(D/2)(2-D)(3-D)},
\end{align}
where we have also employed the relation between Gamma Euler functions $z\Gamma(z)=\Gamma(1+z)$ \cite{AS,21}. 

The above result is still divergent for $D=3$, but we can apply the minimal subtraction method by computing the residue of $I(D)$ in $D=3$:
\begin{equation}
\text{Res}\, I(D)= \lim_{D\rightarrow 3} (3-D) I(D) = -\frac{p}{4\pi^2}.
\end{equation}

Then we can define the regularized integral as:
\begin{equation}
I(D)_r = I(D) - \frac{p\mu^{D-3}}{4\pi^2 (3-D)},
\end{equation} 
where $\mu$ is a mass paramter which allows us to introduce a counter term of the same dimension as $I(D)$.

Explicitly this is:
\begin{align}
I(D)_r = \frac{1}{3-D} \bigg(\frac{2\pi^{D/2}p^{D-2}\Gamma(D-1)\Gamma(4-D)}{2(2\pi)^D\Gamma(D/2)(2-D)}  + \frac{p\mu^{D-3}}{4\pi^2} \bigg).
\end{align} 

By expanding the parenthesis around $D=3$ we obtain a term proportional with $3-D$, that will cancel the divergence. The final result is:
\begin{align}
&&I(D)_r = \frac{p}{8\pi^2}\left[2-\gamma-\ln\left(\frac{p^2}{\pi\mu^2}\right)\right]+\mathcal{O}\left((D-3)^2\right),
\end{align}
where terms $\mathcal{O}\left((D-3)^2\right)$ will vanish for $D=3$. 

We obtain at last the total transition rate in the large expansion limit:
\begin{align}
R_{{ e^{-} \rightarrow W^{-} + \nu}} = \frac{p M_W^2 G_F}{8\pi^2\sqrt{2}(2\pi)^3}\left[2-\gamma-\ln\left(\frac{p^2}{\pi\mu^2}\right)\right],
\end{align}
where we have made use of the relation between the Fermi constant $G_F$ and the coupling constant $g$ \eqref{ctcu}.

For the total transition rate to be well defined, the argument of the logarithm has to be subunitary.

\section{\bf{Conclusions}}

In the present paper we investigated the decays of $W$ bosons and the emission of $W$ bosons by electrons in the  de Sitter spacetime. These elementary processes correspond to the first order of perturbation theory, and can be studied by using the electro-weak charged current interactions in the de Sitter background. We computed the transition rates for the processes in which the $W$ bosons can be produced by a charged leptons, and the decay rates of the $W$ bosons into leptons in the de Sitter spacetime. We also explored the limit of a large expansion factor, when the Hubble parameter becomes much larger than the particle masses. The transition rates in the strong gravitational regime are obtained as finite quantities with the help of the dimensional regularization method, combined with the minimal subtraction method \cite{WF,GHV,IT,GT,HGF}. 

The temporal integrals in our amplitudes contain a dependence on the particles masses and the expansion parameter. In the general case, when the amplitudes depend on the Hubble constant, particles masses and momenta, the integrals that need to be solved for extracting the total probabilities and total transition rates are not solvable analytically. Under these circumstances we look for various relevant limits of our general results, which lead to the study of the Minkowski limit, as well as the high expansion limit, $M_W/\omega \rightarrow 0$, which corresponds to the early Universe. In the limit $M_W/\omega \rightarrow 0$, the momenta integrals that need to be computed in order to obtain the transition rates are also divergent, but unlike the general case, here they can be solved by using well known regularization methods \cite{WF,GHV,IT,GT,HGF}.

We have managed to recover the Minkowski limit of the transition rate for the leptonic decay of a $W^-$ boson, which is equal to the Minkowski transition rate for the same process \cite{rat}. For obtaining this limit we considered the ultra-relativistic case, when the modulus of the momentum is equal to the energy of the particle. This is a remarkable result, and proves that our general findings in the de Sitter Universe can lead to the Minkowski formulas for $\omega=0$.

We also proved that the emission of a $W$ boson by a charged lepton is possible as a perturbative process only in the high expantion conditions of the early Universe, as we obtained that the rate for this process vanishes in the Minkowski limit. Our investigations confirm results obtained by both perturbative \cite{23,rc,24,27,cc,43,44,45,46} and non-perturbative methods \cite{32,33,34,35}, where the conclusions lead to the fact that particle generation is possible only in the early Universe.\\

\section*{\bf{Acknowledgement}}
This work was supported by the European Union - NextGenerationEU through
the grant No. 760079/23.05.2023, funded by the Romanian ministry of
research, innovation and digitalization through Romania’s National
Recovery and Resilience Plan, call no. PNRR-III-C9-2022-I8.

\vspace{0.2cm}
\noindent

\appendix

\section{\bf{Mathematical formulas}}\label{AppendixA}

In this section we provide some of the formulas which we have used throughout this paper.

The relations between exponentials, Hankel functions and Bessel J and Bessel K functions are \cite{AS,21}:
\begin{align}\label{exptok}
e^{-z} = \sqrt{\frac{2 z}{\pi}}K_{1/2}(z);
\end{align}

\begin{align}\label{hankeltok}
H^{(2),(1)}_{\nu} (z) = \pm \frac{2 i}{\pi} e^{\pm i\pi\nu/2}K_{\nu}(\pm i z);
\end{align}

\begin{align}\label{hankeltoj}
H^{(2),(1)}_{\nu} (z) = \frac{J_{-\nu}(z)-e^{\pm i\pi\nu}J_{\nu}(z)}{\mp i \sin{(\pi\nu)}}.
\end{align}

The integral that helps us obtain the transition amplitudes is \cite{AS,21}:
\begin{align}\label{intBesseltoF4}
\int_{0}^{\infty} &dx\, x^{Q-1} J_{\lambda}(ax) J_{\mu}(bx)K_{\nu}(cx) = \frac{2^{Q-2} a^{\lambda}b^{\mu}c^{-Q-\lambda-\mu}}{\Gamma(\lambda + 1)\Gamma(\mu+1)} \nonumber\\
&\,\,\times \Gamma\left(\frac{Q+\lambda+\mu-\nu}{2}\right) \Gamma\left(\frac{Q+\lambda+\mu+\nu}{2}\right)\nonumber\\
&\,\,\times F_{4} \Bigg( \frac{Q+\lambda+\mu-\nu}{2}, \frac{Q+\lambda+\mu+\nu}{2}, \lambda +1, mu+1, -\frac{a^2}{c^2}, -\frac{b^2}{c^2}\Bigg).
\end{align}

The definition of the Appell $F_{4}$ function is given below \cite{AS,21}:
\begin{align}\label{F4series}
F_{4}(\alpha,\beta,\gamma,\gamma';x,y) &= \sum_{m=0}^{\infty} \sum_{n=0}^{\infty} \frac{\Gamma(\alpha+m+n)\Gamma{
		(\gamma)}}{\Gamma(\alpha)\Gamma(\gamma+m)} \frac{\Gamma(\beta + m + n)\Gamma(\gamma')}{\Gamma(\beta)\Gamma(\gamma' +n)}\frac{x^{m}y^{n}}{m!n!}.
\end{align}

The Beta Euler function (which we employ in dimensional regularisation) is:
\begin{align}\label{betaeuler}
B(\alpha,\beta) = \frac{\Gamma(\alpha) \Gamma(\beta)}{\Gamma(\alpha + \beta)} = \int_{0}^{\infty} dy \frac{y^{\alpha - 1}}{(1 + y)^{\alpha + \beta}}.
\end{align}

The Gamma function can be expanded as:
\begin{align}\label{gammaexpansion}
\Gamma(\epsilon - n) = \frac{(-1)^n}{n!} \bigg\{ \frac{1}{\epsilon} + \psi(n+1) \frac{\epsilon}{2} \bigg[\frac{\pi^2}{3} + \psi^2(n+1)
- \psi'(n+1)\bigg] +\mathcal{O}\left(\epsilon^2\right) \bigg\},
\end{align}
where $\psi$ are the Digamma functions:
\begin{align}
\psi(n) &= -\gamma + \sum_{l=1}^{n-1} \frac{1}{l},\\
\psi'(n) &= \frac{\pi^2}{6} - \sum_{l=1}^{n-1} \frac{1}{l^2},
\end{align}
and $\gamma = - 0.5772$ is the Euler constant.

\section{\bf{Computing the amplitude in the Minkowski limit}}\label{AppendixB}

In this section we give the steps to solve the temporal integral contained in amplitude \eqref{amm} from section \hyperref[sec2.4]{2.4}, which helps us recover the Minkowski limit of the transition rate for the process $W^{-} \rightarrow e^{-} + \tilde{\nu}_e$.

For solving this integral we begin with the temporal integral in \eqref{AmpDecayWz} and take the fermion mass equal to zero. First we transform the electron Hankel function into an exponential by using formulas \eqref{hankeltok} and \eqref{exptok}:
\begin{align}\label{limminkwt1}
e^{-\pi K/2}\int_{0}^{\infty} dz \,z \, e^{-ip'z}H_{\frac{1}{2}}^{(2)} (pz)  H_{iK}^{(1)} (Pz) = e^{-\pi K/2}e^{i\pi/4}\sqrt{\frac{2i}{\pi p}} \int_{0}^{\infty} dz \, \sqrt{z} \, e^{-i(p+p')z} H_{iK}^{(1)} (Pz).
\end{align}

Afterwards we transform the exponential function into a Hankel function of the second kind by using the same formulas as before. Furthermore we transform the remaining Hankel function into a Bessel K function by using formula \eqref{hankeltok}:
\begin{equation}\label{limminkwt2}
\frac{-2i}{\pi} \sqrt{\frac{(p+p')}{p}} \int_{0}^{\infty} dz \, z \, H_{\frac{1}{2}}^{(2)} \big((p+p')z\big) K_{iK} (-iPz),
\end{equation}
from which we can write the final form of this integral by transforming the Hankel functions into Bessel J \eqref{hankeltoj}:
\begin{align}\label{kjj}
\frac{-2i}{\pi}\sqrt{\frac{p+p'}{p}}  \int_{0}^{\infty} dz \, z\, K_{iK} (-iPz)\bigg( J_{\frac{1}{2}}\big((p+p')z\big) + i J_{-\frac{1}{2}} \big((p+p')z\big) \bigg).
\end{align}

We mention that factor $-2i/\pi\sqrt{p}$ will be reduced by other terms in amplitude \eqref{amm}, while the term $\sqrt{p+p'}$ will remain moving forward.

In order to solve \eqref{kjj} further we add a small imaginary part to the $W$ boson momentum $P\rightarrow P+i\epsilon$, for the sake of assuring the convergence of the integral. At the end of our computations we will take the limit $\epsilon\rightarrow 0$.

Next we divide \eqref{kjj} into two integrals:
\begin{align}
I=I_1+I_2 = \sqrt{p+p'}  \int_{0}^{\infty} dz \, z \, K_{iK} (-i(P+i \epsilon)z)\bigg( J_{\frac{1}{2}}\big((p+p')z\big) + i J_{-\frac{1}{2}} \big((p+p')z\big) \bigg),
\end{align}
which can be evaluated by using \cite{AS,21}:
\begin{align}\label{a3}
\int_0^{\infty} dz \, z^{-\lambda}J_{\nu}(\alpha z)K_{\mu}(\beta z) &=\frac{\alpha^{\nu}}{2^{\lambda+1}\beta^{\nu-\lambda+1}\Gamma\left(1+\nu\right)}\nonumber\\
&\times\Gamma\left(\frac{\nu + \mu-\lambda+1}{2}\right)\Gamma\left(\frac{\nu-\mu-\lambda+1}{2}\right)\nonumber\\
&\times\,_{2}F_{1}\left(\frac{\nu +\mu -\lambda+1}{2},\frac{\nu-\mu-\lambda+1}{2};\nu+1;-\frac{\alpha^2}{\beta^2}\right),\nonumber\\
&\text{Re}(\alpha\pm i\beta)>0\,,|\text{Re}(\nu-\lambda+1)|>|Re(\mu)|.
\end{align}

The results for $I_1$ and $I_2$ are:
\begin{align}\label{rezult}
I_1(\epsilon) &= \frac{(p+p')\,\Gamma\left(\frac{5}{4}+\frac{iK}{2}\right)\Gamma\left(\frac{5}{4}-\frac{iK}{2}\right)}{\Gamma\left(\frac{3}{2}\right)(\epsilon-iP)^{5/2}} \,_{2}F_{1}\left(\frac{5}{4}-\frac{iK}{2},\frac{5}{4}+\frac{iK}{2};\frac{3}{2};\frac{(p+p')^2}{(P+i\epsilon)^2}\right);\\
I_2(\epsilon) &= \frac{\,\Gamma\left(\frac{3}{4}+\frac{iK}{2}\right)\Gamma\left(\frac{3}{4}-\frac{iK}{2}\right)}{\Gamma\left(\frac{1}{2}\right)(\epsilon-iP)^{3/2}} \,_{2}F_{1}\left(\frac{3}{4}-\frac{iK}{2},\frac{3}{4}+\frac{iK}{2};\frac{1}{2};\frac{(p+p')^2}{(P+i\epsilon)^2}\right).
\end{align}

The above results for the integrals contain both cases of $P=p+p'$ and $P\neq p+p'$. To obtain the result in the case when $P=p+p'$, we transform the Gauss hypergeometric function according to the formula \cite{AS,21}:
\begin{equation}
_{2}F_{1}(a,b;c;z) = (1-z)^{c-a-b} \, _{2}F_{1}(c-a,c-b;c;z).
\end{equation}

Thus we obtain for $I_1(\epsilon)$:
\begin{align}
&_{2}F_{1} \left( \frac{5}{4}-\frac{iK}{2},\frac{5}{4}+\frac{iK}{2}; \frac{3}{2}; \frac{(p+p')^2}{(P+i\epsilon)^2} \right) =\nonumber\\ 
&=\left(1-\frac{(p+p')^2}{(P+i\epsilon)^2}\right)^{-1} 
\,_{2}F_{1} \left( \frac{1}{4}-\frac{iK}{2},\frac{1}{4}+\frac{iK}{2}; \frac{3}{2}; \frac{(p+p')^2}{(P+i\epsilon)^2} \right).
\end{align}

The momentum dependent factor in front of the hypergeometric function can be written as:
\begin{align}
\left(1-\frac{(p+p')^2}{(P+i\epsilon)^2}\right)^{-1}=-\frac{(P+i\epsilon)^2}{P\left(\frac{(p+p')^2+\epsilon^2}{P}-P-2i\epsilon\right)},
\end{align}
while in $I_1(\epsilon)$ we can transform:
\begin{equation}
(\epsilon-iP)^{-5/2}=-e^{i\pi/4}(i\epsilon+P)^{-5/2}.   
\end{equation}

Then the first integral $I_1(\epsilon)$ can be brought to the form:
\begin{align}\label{rezult12}
I_1(\epsilon)&=\frac{e^{i\pi/4}(p+p')}{\left(\frac{(p+p')^2+\epsilon^2}{2P}-\frac{P}{2}-i\epsilon\right)}\frac{\Gamma\left(\frac{5}{4}+\frac{iK}{2}\right)\Gamma\left(\frac{5}{4}-\frac{iK}{2}\right)}{\Gamma\left(\frac{3}{2}\right)(i\epsilon+P)^{1/2}}\nonumber\\
&\times_{2}F_{1}\left(\frac{1}{4}-\frac{iK}{2},\frac{1}{4}+\frac{iK}{2};\frac{3}{2};\frac{(p+p')^2}{(P+i\epsilon)^2}\right).
\end{align}

We can take limit $\epsilon=0$ by using the principal part prescription \cite{AS,21}:
\begin{equation}
\lim_{\epsilon\rightarrow 0}\frac{1}{x-x_0\pm i\epsilon}=Pp\left(\frac{1}{x-x_0}\right)\mp i\pi \delta(x-x_0),
\end{equation}
where the first term of the result is given in equation (\ref{rezult}) and depends on gamma Euler functions and the hypergeometric function. The second term proportional with the delta Dirac function is obtained for the case when the momenta are conserved in this process. 

In our case the limit gives:
\begin{align}
\lim_{\epsilon\rightarrow 0}\frac{1}{\left(\frac{(p+p')^2+\epsilon^2}{2P}-\frac{P^2}{2P}-i\epsilon\right)}&=i\pi\delta\left(\frac{(p+p')^2-P^2}{2P}\right)+Pp\left(\frac{1}{\left(\frac{(p+p')^2}{2P}-\frac{P^2}{2P}\right)}\right).
\end{align}

The final result for $I_1$ when $P=p+p'$ is:
\begin{equation}\label{rezult13}
\lim_{\epsilon\rightarrow 0}I_1|_{P=p+p'} = \frac{i\pi e^{i\pi/4}}{2}\frac{(p+p')}{P^{3/2}}\delta(p+p'-P),
\end{equation}
where we have used relation \cite{AS,21}:
\begin{equation}
_{2}F_{1}(a,b;c;1)=\frac{\Gamma(c)\Gamma(c-a-b)}{\Gamma(c-a)\Gamma(c-b)}
\end{equation}
for rewriting the hypergeometric function in \eqref{rezult12}.

The result for the second integral $I_2$ in the limit $\epsilon\rightarrow 0$ can be obtained by following the same method:
\begin{equation}
\lim_{\epsilon\rightarrow 0}I_2|_{P=p+p'}=\frac{i\pi e^{-i\pi/4}}{2 P^{1/2}}\delta(p+p'-P).
\end{equation}

In the case when  $p+p'\neq P$ the result for integrals $I_1 $and $I_2$ can be expressed in terms of unit step functions as such:
\begin{align}
I_1 + I_2 |_{P\neq p+p'} &= \Theta(P-p-p')\left[f_1\left(\frac{(p+p')}{P}\right)+f_2\left(\frac{(p+p')}{P}\right)\right]\nonumber\\
&+\Theta(p+p'-P)\left[f_1'\left(\frac{P}{(p+p')}\right)+f_2'\left(\frac{P}{(p+p')}\right)\right],
\end{align}
where functions $f_1 $and $f_2$ are defined as follows:
\begin{align}
f_1 \left(\frac{(p+p')}{P}\right) &= \frac{(p+p')\,\Gamma\left(\frac{5}{4}+\frac{iK}{2}\right)\Gamma\left(\frac{5}{4}-\frac{iK}{2}\right)}{\Gamma\left(\frac{3}{2}\right)
	(-iP)^{5/2}}\nonumber\\
&\times\,_{2}F_{1}\left(\frac{5}{4}-\frac{iK}{2},\frac{5}{4}+\frac{iK}{2};\frac{3}{2};\frac{(p+p')^2}{P^2}\right);\\
f_2 \left(\frac{(p+p')}{P}\right) &= \frac{\,\Gamma\left(\frac{3}{4}+\frac{iK}{2}\right)\Gamma\left(\frac{3}{4}-\frac{iK}{2}\right)}{\Gamma\left(\frac{1}{2}\right)
	(-iP)^{3/2}}\nonumber\\
&\times\,_{2}F_{1}\left(\frac{3}{4}-\frac{iK}{2},\frac{3}{4}+\frac{iK}{2};\frac{1}{2};\frac{(p+p')^2}{P^2}\right);
\end{align}
while functions $f_1'$ and $f_2'$ are:
\begin{align}
f_1 \left(\frac{P}{(p+p')}\right) &=\frac{\sqrt{(p+p')}e^{\pi K/2}e^{i\pi/4}}{P^{-iK}\cosh{(\pi K})} \frac{\Gamma\left(\frac{5}{4}+\frac{iK}{2}\right)\Gamma\left(\frac{3}{4}+\frac{iK}{2}\right)}{\Gamma\left(1+iK\right)(p+p')^{iK+2}}\nonumber\\
&\times\,_{2}F_{1}\left(\frac{5}{4}+\frac{iK}{2},\frac{3}{4}+\frac{iK}{2};1+iK;\frac{P^2}{(p+p')^2}\right);\\
f_2'  \left(\frac{P}{(p+p')}\right) &= -\frac{\sqrt{(p+p')}e^{\pi K/2}e^{i\pi/4}}{P^{iK} \cosh{(\pi K)}}\frac{\Gamma\left(\frac{5}{4}-\frac{iK}{2}\right)\Gamma\left(\frac{3}{4}-\frac{iK}{2}\right)}{\Gamma\left(1-iK\right)(p+p')^{2-iK}}\nonumber\\
&\times\,_{2}F_{1}\left(\frac{5}{4}-\frac{iK}{2},\frac{3}{4}-\frac{iK}{2};1-iK;\frac{P^2}{(p+p')^2}\right).
\end{align}

All these functions vanish in the Minkowski limit, when $M_W/\omega$ becomes very large. As a result the only contribution to the amplitude of $W^-$ decay in this limit comes from the terms proportional with the delta Dirac function $\delta(p+p'-P)$.


\begin{thebibliography}{99}
	\bibitem{1}
	C. W. Misner, K. S. Thorne and J. A. Wheleer, {\em Gravitation}
	(W. H. Freeman and Company New York, 1973).
	\bibitem{PR1}
	A. Proca, {\em J. Phys. Radium} \textbf{7},
	347–353 (1936).
	\bibitem{pr2}
	A. Proca, {\em C. R. Acad. Sci. Paris}, \textbf{202}, 1366 (1936).
	\bibitem{pr3}
	A. Proca, {\em C. R. Acad. Sci. Paris} \textbf{202}, 1490
	(1936).
	\bibitem{pr4}
	A. Proca, {\em  J. Phys. Radium} \textbf{9}, 61 (1938).
	\bibitem{2}
	Ion I. Cot\u{a}escu, {\em Gen.Rel.Grav.} \textbf{42},861-876,2010.
	\bibitem{w1}
	S. Weinberg, The First Three Minutes: A Modern View of the Origin of the Universe (Basic Books,New York, 1977).
	\bibitem{w2}
	S. Weinberg, {\em Phys. Scr.} \textbf{21}, 773 (1979).
	\bibitem{3}
	S. Weinberg, {\em Phys. Rev. Lett.} \textbf{19}, 1264 (1967).
	\bibitem{4}
	S. Weinberg, {\em Phys. Rev. Lett.} \textbf{27}, 1688 (1971).
	\bibitem{5}
	S. Weinberg, {\em Phys. Rev. D} \textbf{5}, 1412 (1972).
	\bibitem{6}
	S. Weinberg, {\em Phys. Rev. D} \textbf{7}, 1068 (1973).
	\bibitem{7}
	S. Weinberg, {\em Phys. Rev. D} \textbf{8}, 605 (1973).
	\bibitem{8}
	S. Weinberg, {\em Rev. Mod. Phys.} \textbf{46}, 255 (1974).
	\bibitem{9}
	S. L. Glasshow and S. Weinberg, {\em Phys. Rev. D} \textbf{15}, 1958 (1977).
	\bibitem{10}
	J. D. Bjorken, K. Lane and S. Weinberg, {\em Phys. Rev. D} \textbf{5}, 1474 (1977).
	\bibitem{11}
	B. W. Lee and S. Weinberg, {\em Phys. Rev. D} \textbf{38}, 1237 (1977).
	\bibitem{cr}
	C. Rubbia, {\em Rev. Mod. Phys.} \textbf{57}, 699 (1985).
	\bibitem{12}
	S. Weinberg, {\em The Quantum Theory of Fields}  (Cambridge University Press, Cambridge, 1995).
	\bibitem{15}
	J. Lankinen and I. Vilja, {\em Phys. Rev. D} \textbf{96}, 105026-1 (2017).
	\bibitem{17}
	N. D. Birrel and P. C. W. Davies,  {\em Quantum Fields in Curved Space} (Cambridge University Press, Cambridge 1982).
	\bibitem{18}
	N. D. Birrel, P. C. W. Davies and L. H. Ford, {\em J. Phys. A} \textbf{13}, 961 (1980).
	\bibitem{19}
	S. Drell and J. D. Bjorken, {\em Relativistic Quantum Fields} (Mc Graw-Hill Book Co., New York 1965).
	\bibitem{20}
	L. Landau and E. M. Lifsit, {\em Theorie Quantique Relativiste} (Mir Moscou 1972).
	\bibitem{AS}
	M. Abramowitz and I. A. Stegun, {\em Handbook of Mathematical Functions} (Dover, New York, 1964).
	\bibitem{21}
	I. S. Gradshteyn and I. M. Ryzhik {\em Table of integrals, series and products} (Academic Press, 2007).
	\bibitem{PT}
	T. Prokopec, N. C. Tsamis and R. P. Woodard, {\em AnnalsPhys.} \textbf{323}, 1324-1360,(2008).
	\bibitem{WT}
	N. C. Tsamis and R. P. Woodard, {\em J.Math.Phys.} \textbf{48}, 052306, (2007).
	\bibitem{PC}
	J. F. Koksma, T. Prokopec, {\em Class.Quant.Grav.} \textbf{26}, 125003, (2009).
	\bibitem{CRR}
	P. Candelas and D. J . Raine, {\em Phys. Rev. D} \textbf{12}, 965, (1975).
	\bibitem{COT}
	Ion I. Cot\u{a}escu, {\em Eur. Phys. J. C} \textbf{78:769}, (2018).
	\bibitem{WF}
	M.E. Fisher and K.G. Wilson, {\em Phys. Rev. Lett.} \textbf{28}, 240 (1972);
	K.G. Wilson, {\em Phys. Rev. D} \textbf{7}, 2911 (1973).
	\bibitem{GHV}
	G.’t Hooft and M. Veltman, {\em Nucl. Phys. B} \textbf{44}, 189 (1972).
	\bibitem{IT}
	C.G. Bollini, J.J. Giambiagi, {\em Nuovo Cimento B} \textbf{12}, 20 (1972).
	\bibitem{GT}
	G.’t Hooft, {\em Nucl. Phys. B} \textbf{61}, 455 (1973).
	\bibitem{PV}
	W. Pauli and F. Villars, {\em Rev. Mod. Phys.} \textbf{21}, 434 (1949).
	\bibitem{HGF}
	H. Kleinert and V. Schulte-Frohlinde, {\em Critical properties of $\phi^4$-theories} (World Scientific 2001).
	\bibitem{22}
	Ion I. Cot\u{a}escu, {\em Phys. Rev. D} \textbf{65}, 084008 (2002).
	\bibitem{LL}
	J. Lankinen and I. Vilja, {\em Phys. Rev. D} \textbf{96}, 105026-1 (2017)
	\bibitem{LL1}
	J. Lankinen, J. Malmi and I. Vilja, {\em Eur. Phys. J. C} \textbf{80:502}, (2020).
	\bibitem{23}
	Crucean Cosmin, {\em Phys. Rev. D} \textbf{85}, 084036 (2012).
	\bibitem{rc}
	C. Crucean and R. Racoceanu, {\em Int. J. Mod. Phys. A} \textbf{23}, (2008).
	\bibitem{24}
	Ion I. Cot\u{a}escu, C. Crucean, {\em Phys. Rev. D} \textbf{87}, 044016 (2013).
	\bibitem{25}
	Ion I. Cot\u{a}escu, C. Crucean, {\em Progress of Theor. Phys.} \textbf{124}, 1051 (2010).
	\bibitem{26}
	C. Crucean and M. A. B\u{a}loi {\em Phys. Rev. D} \textbf{93}, 044070 (2016).
	\bibitem{27}
	C. Crucean, {\em Mod. Phys. Lett. A} \textbf{22}, 2573 (2007).
	\bibitem{28}
	C. Crucean and M. A. B\u aloi, {\em Int. J. Mod. Phys. A} \textbf{30}, 1550088 (2015).
	\bibitem{29}
	Ion I. Cotaescu, R. Racoceanu, Radu and C. Crucean, {\em Mod. Phys. Lett. A} \textbf{21}, 1313 (2006).
	\bibitem{30}
	Ion I. Cot\u{a}escu, C. Crucean, {\em Int. J. Mod. Phys. A} \textbf{23}, 3707 (2008).
	\bibitem{31}
	M. A. B\u{a}loi, {\em Mod. Phys. Lett. A} \textbf{29}, 1450138 (2014).
	\bibitem{b1}
	M. A. B\u{a}loi,{\em Int. J. Mod. Phys. A} \textbf{31}, 1650081 (2016).
	\bibitem{b2}
	M. A. B\u{a}loi, C. Crucean and D. Popescu, {\em Eur. Phys. J. C} \textbf{78:398}, (2018).
	\bibitem{cc}
	C. Crucean, {\em Eur. Phys. J. C} \textbf{79:483}, (2019).
	\bibitem{32}
	E. Schr\" odinger, {\em Physica} \textbf{6}, 899 (1939).
	\bibitem{33}
	L. Parker, {\em Phys. Rev. Lett.} \textbf{21}, 562 (1968).
	\bibitem{34}
	L. Parker, {\em Phys. Rev.} \textbf{183}, 1057 (1969).
	\bibitem{35}
	L. Parker, {\em Phys. Rev. D} \textbf{3}, 346 (1971).
	\bibitem{cpc}
	Ion I. Cot\u{a}escu, D. Popescu, {\em Chinese Phys. C} \textbf{44}, (2020).
	\bibitem{37}
	Y. Ema, K. Nakayama and Y. Tang, {\em JHEP} \textbf{60} (2019).
	\bibitem{rat}
	W. N. Cottingham, D. A. Greenwood, {\em An introduction to standard model of particle physics}, (Cambridge University Press, Cambridge 2007).
	\bibitem{rfv}
	Ion I. Cot\u{a}escu, {\em Eur. Phys. J. C} \textbf{79:696}, (2019).
	\bibitem{rfv1}
	Ion I. Cot\u{a}escu, {\em Eur. Phys. J. C} \textbf{80:535}, (2020).
	\bibitem{38}
	B. Allen and T. Jacobson, {\em Commun. Math. Phys.} \textbf{103}, 669-692 (1986).
	\bibitem{40}
	T. Prokopec, O. Törnkvist, R.P. Woodard, {\em Phys. Rev. Lett.} \textbf{89}, 101301 (2002).
	\bibitem{41}
	T. Prokopec, O. Törnkvist, R.P. Woodard, {\em Annals of Physics} \textbf{303(2)}, 251-274 (2002).
	\bibitem{43}
	D. Dumitrele ; M. A. B\u aloi; C. Crucean,  {\em Eur. Phys. J. C} \textbf{83:738}, 2023.
	\bibitem{44}
	C. Crucean ; A. D. Fodor, {\em Eur. Phys. J. C} \textbf{83:929}, 2023.
	\bibitem{45}
	C. Crucean, D. Dumitrele {\em Eur. Phys. J. C} \textbf{84:855}, 2024.
	\bibitem{46}
	M. A. B\u aloi, C. Crucean, {\em Int.J.Mod. Phys. A} \textbf{32}, 1750208, (2017);
	\bibitem{47}
	M. A. B\u aloi, C. Crucean, D. Popescu, {\em Eur. Phys. J. C} \textbf{78:398}, 2018.
	\bibitem{auluk}
	S. K. H. Auluck, {\em The Mathematica Journal} \textbf{14}, (2012).
\end{thebibliography}
\end{document}